%% file: Main.tex
\documentclass[12pt]{elsarticle}
\input{Preamble.tex}

\begin{document}
\doublespacing

\begin{frontmatter}
\title{\large \textbf {An assessment of the \j-integral test for a metallic foam}}

\author[cam]{H.~C.~Tankasala} \ead{hct30@cam.ac.uk}
\author[cam]{T.~Li} 
\author[cam]{P.~E.~Seiler} 
\author[cam]{V.~S.~Deshpande} 
\author[cam]{N.~A.~Fleck\corref{cor}} \ead{naf1@eng.cam.ac.uk}
\cortext[cor]{Corresponding author}
\address[cam]{Department of Engineering, University of Cambridge,Cambridge CB2 1PZ, United Kingdom}

\begin{abstract}
An assessment is made of the \j-integral test procedure for initial crack growth in an open-cell aluminium alloy foam by combining finite element (FE) simulations with experiment. It is found experimentally that a zone of randomly failed struts develops ahead of the primary crack tip, and is comparable in size to that of the plastic zone. Hence, a crack tip \j-field is absent at the initiation of crack growth from the primary crack tip. This implies that the measured \j$_{\rm IC}$ value and the  \j\xspace versus crack extension $\Delta{a}$ curve cannot be treated as material properties despite the fact that the specimen size meets the usual criteria for \j\xspace validity. The toughness tests were performed on a single-edge notched bend specimen, and crack extension was measured by the direct current potential drop method, by digital image correlation and by X-ray computed tomography. The crack growth resistance of the foam is associated with two distinct zones of plastic dissipation: (i) a bulk plastic zone emanating from the crack tip (containing a cluster of randomly failed struts), and (ii) a crack bridging zone behind the advancing crack tip. The applicability of a cohesive zone model to predict the fracture response is explored for the observed case of large scale bridging. To do so, FE simulations are performed by replacing the discrete lattice of the open-cell metallic foam by a compressible, elastic-plastic hardening solid while the fracture process  zone in the foam is represented by a cohesive zone, as characterised by a tensile traction versus separation law. A detailed comparison of the cohesive zone model with experimental observations reveals that it is possible to capture the load versus displacement response but not the details of the fracture process zone using a single set of process zone parameters. 
\end{abstract}


\end{frontmatter}

%------------------------------------------------------------------------------------------------------
% INTRODUCTION AND LIT REVIEW
%------------------------------------------------------------------------------------------------------
\section{Introduction}   
Metallic foams enjoy increasing application in sandwich panels for lightweight structural components, in energy absorption systems for protection from impacts, in heat sinks for electronic devices and in acoustic insulation, \textit{inter alia}. The focus of this study is the fracture response of open-cell aluminium alloy foams.

The standard method for measuring the fracture toughness and crack growth resistance of ductile materials, including metallic foams, is the \j$-$integral procedure as outlined in the ASTM standard E1820~\citep{e1820}, see for example \citet{Jelitto2019} for a review of fracture toughness tests on porous materials. The ability of the \j$-$integral to characterise crack initiation and crack growth is predicated upon the existence of a near-tip \j$-$field, such as the HRR field in a fully dense elasto-plastic solid. This \j$-$field must encompass the fracture process zone (FPZ) at the crack tip, as sketched in \cref{fig:sketch_ssb_lsb}(a). Recall that the FPZ is on the order of the crack tip opening displacement for a fully dense metallic solid. At issue is whether the FPZ much exceeds the cell size $\ell$ of a metallic foam, particularly when failure is stochastic in nature and involves the failure of struts in a zone ahead of the crack tip. 

Previous studies of the fracture of metallic foams assume that a crack bridging zone develops only after advance of the main crack tip, with successive strut-by-strut failure occurring at the crack tip \citep{McCullough1999,Kashef2010,Combaz2010}. However, this assumption needs to be validated by experiment. It may be that a large FPZ develops ahead of the main crack tip in metallic foams, with a stochastic distribution of strut failure in existence before the main crack tip advances. If this FPZ is comparable in size to the plastic zone size, as illustrated in \cref{fig:sketch_ssb_lsb}(b), then no \j$-$field exists, regardless of the overall size of the specimen and of the initial crack length. The relative size of FPZ to plastic zone is unclear from the literature on metallic foams, and is measured in the present study.

\citet{Combaz2010} have performed \j-tests on compact tension specimens made from aluminium-replicated open-cell foams. Their open-cell foams have a small and relatively uniform cell diameter ($\approx \SI{400}{\micro m}$) and are made from highly ductile pure aluminium. They observed that strut fracture was stochastic in a zone ahead of the crack tip, and the resulting crack extension is accompanied by the development of a substantial  crack-bridging zone. For example, their bend specimen had a net section ligament of length $10$\,mm, ahead of the pre-crack, at the start of the toughness test, and they report a large scale bridging zone of length $5$\,mm behind the crack tip after crack advance on the order of $7$\,mm. Hence, it is unlikely that a \j$-$field surrounds the advancing crack tip. Similar observations on the existence of a bridging zone (of size 4 to 8 cells) in the wake of an advancing crack have been reported for the case of open-cell aluminium alloy foams \citep{McCullough1999}, open-cell titanium alloy foams \citep{Kashef2010}, and closed-cell aluminium alloy foams \citep{Olurin2000,Motz2001}. These experimental studies all followed the ASTM standard E1820~\citep{e1820} to measure the crack growth resistance from the fracture response of a bend specimen, and each met the size criteria as demanded by the standard. A significant $R$-curve behaviour is reported in each of these studies due to the presence of crack bridging ligaments behind the advancing crack tip.  \citet{Masta2017} have used the same ASTM procedure for measuring the crack growth resistance of 3D octet-truss lattices. They observed that the fracture toughness scales linearly with the square root of the cell size and linearly with the lattice relative density. However, in contrast to the fracture process observed in foams, the struts of the lattice failed cell by cell at the crack tip, without the formation of a recognisable bridging zone. 
 
\subsection{Prediction of crack growth in metallic foams}
The cohesive zone approach is a useful predictive tool for modelling crack growth in the presence of bulk plasticity as observed in metallic foams. In this approach, the details of the crack tip failure mechanism are not stated explicitly. Rather, a cohesive zone, as defined by a tensile traction versus separation response, idealises the FPZ at the crack tip and is embedded within an elasto-plastic continuum. The cohesive zone approach has the virtue that it can be used to model both small scale and large scale bridging \citep{Sorensen1998, Zok1990}. 

{The discrete foam can be idealized as a compressible continuum when the length scales (associated with the geometry and loading) are much larger than the cell size \citep{Deshpande2000}. Size effects are exhibited when the structural length scale is comparable to the cell size (see e.g. \citep{Tekoglu2008}), and the classical continuum theory must then be extended to include a characteristic length scale in the constitutive description, or the foam microstructure has to be explicitly modeled. Higher-order continuum theories have been used by \citet{Dillard2006} and \citet{Tekoglu2011} to study the strain fields around a hole in a cellular solid. These theories were able to capture some of the observed cell size effects that were observed in the experiments of \citet{Dillard2006}. Discrete micromechanical models of a foam have also been used to study the strain fields around a sharp notch \citep{Andrews2001, Onck2001}.} Onck and co-workers~\citep{Mangipudi2011a, Mangipudi2011b} have explored numerically the fracture of a discrete Voronoi structure under remote tension, and made from a strain-hardening solid. Each cell wall was modelled as an array of beam elements, and damage in the cell wall material was represented by a linear softening stress versus displacement response of the beam element beyond a critical value of stress. They found that the cell walls fail at random locations within the structure. Clusters of 2 to 3 failed struts eventually coalesce to form a single macroscopic crack which leads to  final failure. Predictions  have also been made for the case of regular hexagonal honeycombs in the elastic-brittle regime \citep{Seiler2019a} and in the creep ductile regime \citep{Seiler2019b} by assuming that each cell wall comprises several continuum elements in order to capture the scatter in the strut geometry (i.e. thickness and Plateau border radius). The failure of each strut was modelled by a linear softening constitutive response. The critical cluster size of the failed struts was found to be only a cell size for the elastic-brittle lattice and about half the width of the specimen for lattices in the creep ductile regime. These micromechanical models are able to explain the observed tensile ductility of foams, but their use is limited to small specimens containing only a few unit cells due to the associated computational cost.

In the present study, a cohesive zone approach is used to predict the fracture response of foams. The cohesive zone parameters representing the fracture process are deduced from the measured load versus displacement response of pre-cracked bend specimens. Predictions are then made for crack growth initiation and growth in metallic foams. The feasibility of the cohesive zone approach to model the fracture response is thereby assessed. We begin by reviewing the use of cohesive zone models in non-linear fracture mechanics. 

\subsection{Cohesive zone models}
\citet{Tvergaard1992} considered crack growth at the tip of a long crack within an elasto-plastic solid of yield strength $\sigma_{\rm Y}$ and selected levels of strain hardening. The crack growth resistance curve (or $R-$curve) was predicted by placing a tensile cohesive zone of peak strength $\hat\sigma$ and of toughness $\Gamma_0$ (equal to the area under traction versus separation curve) along the crack plane. The initiation fracture toughness $K_0$ for the onset of crack growth follows directly from the Irwin relation $E\Gamma_0 = \left(1-\nu^2 \right)K_0^2$ in terms of the Young{\textquotesingle}s modulus $E$ and Poisson ratio $\nu$ of the bulk solid. For the choice $\hat\sigma \leq 3\sigma_{\rm Y}$, the cohesive zone strength $\hat\sigma$ is too small to activate plasticity within the surrounding bulk solid, resulting in a flat $R-$curve\footnote{The factor of 3 closely relates to the Prandtl solution for peak traction ahead of the crack tip in an elastic perfectly-plastic incompressible solid.}. In contrast, a rising $R-$curve was observed for  $\hat\sigma > 3\sigma_{\rm Y}$ with the elevation in crack growth resistance extremely sensitive to the value of $\hat \sigma/\sigma_{\rm Y}$. For example, for a solid with mild strain-hardening, the steady state value of fracture toughness $K_{\rm SS}$ (at large crack extensions) increases from $2.2 K_0$ when $\hat\sigma =3.5\sigma_{\rm Y}$ to $4 K_0$ when $\hat\sigma =3.75\sigma_{\rm Y}$. \citet{Chen2001} made similar predictions for the $R-$curve in an elastic-plastic compressible solid. They found that $K_{\rm SS}/K_0$ equals unity for $\hat\sigma \approx \sigma_{\rm Y}$ and increases dramatically with increasing $\hat\sigma/ \sigma_{\rm Y}>1$. This reduction in critical value of $\hat\sigma$ from $3\sigma_{\rm Y}$ to $\sigma_{\rm Y}$ is a direct consequence of the reduction in plastic constraint at the crack tip in a compressible solid. It is apparent from the studies of \citet{Tvergaard1992} and of \citet{Chen2001} that the solid must possess sufficient strain hardening in order to attain the prescribed magnitude of peak traction $\hat\sigma$, and for crack growth to ensue.

It is emphasised that the accuracy of the cohesive zone approach is sensitive to the specification of the cohesive traction versus separation law that describes the fracture process. In general, the cohesive law has 3 main features: (i) a peak value of traction, \shat, (ii) a softening curve associated with the tensile fracture process, and (iii) a toughness, \gc (as defined by the area under the traction versus separation curve). Inverse analysis methods have been used extensively in the literature to determine the precise form of the cohesive law; see \citet{Elices2002} for a review of these methods. They include a standard \textit{J-}integral approach for a crack in a double cantilever beam specimen wherein the cohesive traction is computed as the derivative of the \textit{J-}integral with respect to the crack tip  opening displacement. However, it is challenging to deduce the cohesive zone law from a single geometry and then use it to predict the collapse response of a different geometry or loading condition. 

\subsection{Scope of study}
The aim of this study is to attempt to measure the toughness of an open-cell aluminium alloy foam, and to interpret the toughness in terms of the underlying microstructure. The \j$-$integral test method, as specified in ASTM E1820, is adopted for single-edge-notch-bend (SENB) specimens. The direct current potential drop (DCPD) method is used to infer the degree of crack growth, and Digital Image Correlation (DIC) is used to measure the plastic zone size. The fracture process zone is observed by 3D X-ray computed tomographic (XCT) reconstructions of the specimen, at regular intervals during the test. Thus, the existence of a near tip \j-field is investigated.

A cohesive zone, embedded in a compressible, elastic-plastic hardening solid, is used to predict the fracture response of the foam. The cohesive zone parameters, peak strength and toughness, are deduced from a goodness-of-fit measure between the measured and predicted load versus displacement response of the SENB specimens. The ability of the cohesive zone model to predict the fracture response of the foam is thereby explored. 

%--------------------------------------------------------------------------------------------------- EXPERIMENTS
%----------------------------------------------------------------------------------------------------
\section{Experimental investigation}  \label{sec:experimental_program}
\subsection{Material specification and experimental programme}
Flat panels of open-cell Al6101-T6 foam\footnote{Manufactured by ERG Aerospace Corporation, Oakland, California, USA. Refer to \citet{Ashby2000} for details on the manufacturing process.}, of dimension \SI{300}{mm} x \SI{300}{mm} x \SI{26.4}{mm}, were obtained in two (nominal) relative densities, $\overline\rho=6.6\%$ and $\overline\rho=9.6\%$. The foam microstructure was interrogated by X-ray computed tomography, see \cref{fig:3D_RelDen0066,fig:3D_RelDen0096}. For both relative densities of foam, the mean cell edge length is $\ell = \SI{1.96}{mm}$. % and the mean cell size is \SI{3}{mm} to \SI{5}{mm}. 

The foam panels were machined into the following 4 geometries: (i) dogbone-shaped specimen, as shown in \cref{fig:specimens}(a), for measuring the tensile response of the foam, (ii) compression test specimen, as shown in \cref{fig:specimens}(b), (iii) single edge notched bend (SENB) specimen for measuring toughness, see \cref{fig:specimens}(c), and (iv) bend specimen without a pre-notch in order to assist in the validation of the foam constitutive model as used in the FE simulations. Local variations in the foam microstructure are accompanied by a variation in the relative density $\overline\rho$ from specimen to specimen; consequently, the precise value of $\overline\rho$ for each test specimen was measured by weighing with a digital balance. 

Uniaxial tension tests were performed on the dogbone-shaped specimens of \cref{fig:specimens}(a)  using a  screw-driven test machine at a constant displacement rate of \SI{0.01}{mm/s}. The cross-sectional area of the uniform section was \SI{30}{mm} x \SI{26.4}{mm}, and the total length of the specimen (excluding the end tabs) was \SI{100}{mm}. The foaming rise direction of all specimens is the out-of-plane $z-$direction, see \cref{fig:specimens}(a). The end tabs of the specimen were filled with epoxy for local reinforcement. Circular holes of diameter \SI{10}{mm} were drilled at the centre of the epoxy-filled tabs to facilitate pin-loading of the specimen in the axial $y-$direction. Four repeat tests  were performed for each value of $\overline\rho$. The load was measured by the load cell clamped to the stationary platen of the rig, and the average axial strain in the specimen was measured by an extensometer of gauge length $L_{g}=$ \SI{50}{mm}  attached at mid-length of the specimen, see \cref{fig:specimens}(a). 

Uniaxial compression tests were performed on specimens of geometry shown in \cref{fig:specimens}(b) to measure the in-plane response in the $x$ and $y$ directions. The shortening of the specimen was used to define the compressive strain in the specimen. Both platens were lubricated with PTFE spray to reduce friction. Four repeat tests for each value of $\overline\rho$ were performed at a constant displacement rate of \SI{0.01}{mm/s}. The plastic \poisson ratio $\nu^{\rm P}$ of the foam was defined as the negative ratio of the lateral logarithmic strain to the axial logarithmic strain. The value of $\nu^{\rm P}$ was determined from the uniaxial compression test using Digital Image Correlation (DIC) by tracking  two facet points, one on each lateral edge of the specimen (with respect to the loading direction), during the course of the test. 

\subsection{\textit{J-}test procedure}
The crack growth resistance of the aluminium alloy foam was determined from the single edge notch bend (SENB) specimens, following the \j-integral test procedure as outlined in ASTM standard E1820 \citep{e1820}. The direct current potential drop (DCPD) method was used to infer crack extension in the foam specimen. Details of the specimen geometry and test apparatus are given in \cref{fig:specimens}(c). The test procedure is summarised below.

\subsubsection{Specimen size and loading}
The length of pre-crack, $a_0$, and of ligament, $W-a_0$, were chosen to satisfy the ASTM requirements for a valid $J_{\rm IC}$ test (for fully dense metals) as given by:
\begin{equation}
    \left(a_0, W-a_0\right) \ge 25\, \delta_{\rm Tc} =  25 \,\dfrac{J_{\rm Q}}{\sigma_{\rm Y}}
    \label{eqn:sizecriterion}
\end{equation}   
where $\delta_{\rm Tc}$ is the critical crack tip opening displacement, $J_{\rm Q}$ is the provisional value of initiation toughness from the single specimen technique prescribed in \citet{e1820}, and $\sigma_{\rm Y}$ is the tensile yield strength of the foam as measured from the uniaxial tensile test. The size criterion of \cref{eqn:sizecriterion} demands that $\left(a_0, W-a_0\right) > 9$ mm for both $\overline\rho=6.6\%$ and $\overline\rho=9.6\%$. Accordingly, we chose $a_0=$ \SI{20}{mm} and $W=$ \SI{50}{mm} for all SENB specimens tested in this study. The specimens were of span $S=200$ mm between the support rollers and of thickness $B=\SI{26.4}{mm}$. The pre-crack was machined using a fine blade of thickness  $\SI{300}{\micro m}$ which is much less than the cell size of $\ell=$ \SI{1.96}{mm}. 

Specimens were loaded in three-point bending via rollers of diameter \SI{13}{mm}, using a screw-driven testing machine at a constant displacement rate of \SI{0.01}{mm/s}. These rollers ensured that the specimen was electrically insulated from the test machine to facilitate accurate potential drop measurements. The load $P$ was measured via a \SI{2}{kN} load cell clamped to the stationary platen of the rig. The crosshead displacement $v$ was tracked by a non-contact laser extensometer in addition to the DIC instrumentation. Prior to the start of the test, the specimens were spray-painted by black paint in order to enhance the speckle contrast of the DIC images. A single camera of the GOM system\footnote{GOM ARAMIS 12M system, maximum resolution: $4096 \times 3072$ pixels, \SI{100}{mm} lens} was used to track a facet of size $20\times 20$ pixels on the foam directly  beneath the point of contact with the loading roller. An array of additional facets were placed (by the DIC post-processing software) at suitable locations on the surface of the foam specimen to enable the measurement of crack mouth and crack tip opening displacements, and to determine the plastic zone size.  

\subsubsection{Measurement of crack extension by direct current potential drop method}
A constant current of \SI{1}{A} was applied between the ends of the specimen, see  \cref{fig:specimens}(c). The resulting voltage across the crack mouth was measured via soldered probe wires of diameter $0.2$ mm. The location of these probe wires is included in \cref{fig:specimens}(c), and the voltage across the probe wires was recorded by a data logger.  The relationship between the potential drop across probe wires and the crack length was obtained using the electrical analogue method as described by \citet{Smith1974}. Write $V_{\rm o}$ as the voltage across the probe wires for a normalised crack length $a/W$ equal to $a_0/W=0.4$ for the SENB specimen. Then, the relationship between the normalised crack length $a/W$ and the normalised voltage $V/V_0$ was obtained from regression fitting of the data from 3 repeat tests of the electrical analogue to give 
\begin{equation}
    \dfrac{a}{W}=  -0.01478 + 0.50116 \left( \dfrac{V}{V_{\rm o}} \right) - 0.0894 \left( \dfrac{V}{V_{\rm o}} \right)^2 + 0.00557 \left( \dfrac{V}{V_{\rm o}} \right)^3 
\label{eqn:v_calibration}
\end{equation}
{\rewrite{The voltage drop $V/V_0$ increases with increasing crack length $a/W$  according to \cref{eqn:v_calibration}. The increase in voltage (and resistivity) with distributed strut failure has been noted previously by \citet{Amsterdam2008} for the same composition of aluminium alloy foam tested under uniaxial tension and without a pre-crack.}}

Unloading compliance techniques were used to verify the accuracy of the potential drop method in the toughness  tests. The specimens were unloaded by 10\% of the current load, and then re-loaded at regular intervals during the test. The DIC technique was used to track the crack mouth opening displacement and the load-line displacement during the toughness tests. 
  
\subsection{Damage visualisation using X-ray computed tomography}\label{sec:damage-ct}
Three-dimensional computer-assisted X-ray tomography (XCT) was used to map the 3D fracture pattern of failed struts in the foam at selected stages of crack extension $\Delta{a_{\rm PD}}$, as inferred from the potential drop method. The toughness test was interrupted at selected values of $\Delta{a_{\rm PD}}$ and the crack was held open by wedging a PMMA sheet between the crack faces prior to unloading of the specimen in the test machine. High resolution 3D XCT scans of the specimen were obtained by placing the region of interest (in the vicinity of the crack tip) in the detector field-of-view; the typical voxel size of the scans was $\SI{30}{\micro m}$. A series of post-processing steps were carried out in order to identify the broken struts and thereby compute the fracture process zone. The details are as follows. 
First, the 2-dimensional grey-scaled images from the XCT were stacked in a series of thin slices of height $\Delta W = \SI{1.5}{mm}$, length $L=\SI{20}{mm}$, and depth $B = \SI{26.4}{mm}$ into the page, as sketched in \cref{fig:specimens}(d). Each slice was binarised to separate an intact aluminium strut (white) from air (black) by following the method as described in \cite{Otsu1979}. The binarised images were cleaned by removing islands of single pixels or pixels with up to 4 neighbours. Next, the cleaned 3D image of each slice was projected onto a plane by assigning specific grey-scale value to each pixel depending upon its location in the slice. This technique enables the visual identification of the failed struts as well as their location. Finally, an image processing software\footnote{\textit{Fiji} software, \url{https://fiji.sc}} was used to catalogue the number and the $(x, y, z)$ coordinates of the mid-point of each failed strut.

\section{Material characterisation} \label{sec:material_charac}
The in-plane tensile and compressive responses of Al alloy foam of relative density $\overline\rho=6.6\%$ and 9.6\% are  shown in \cref{fig:exp_tension_compression} in terms of the nominal stress $\sigma$ versus nominal strain $\varepsilon$. Data are shown for 4 nominally identical specimens for each relative density, with the solid line corresponding to the mean response and the shaded region representing the scatter in data. The initial response in both uniaxial tension and compression involves elastic bending of the cell walls. In  uniaxial compression, yield occurs at a constant plateau stress prior to significant hardening at a nominal strain beyond $30\%$. In contrast, under remote tensile loading, a few cell walls within a narrow band in the gauge section undergo tensile failure almost immediately after yield, and this leads to a softening post-yield response. The foam ductility is between $2\%$ and $5\%$ depending upon relative density, see \cref{fig:exp_tension_compression}. The out-of-plane compressive strength was higher than the in-plane compressive strength by approximately $10\%$ for both values of relative density. This mild anisotropy is attributed to the foaming process and is ignored in the remainder of the study. The ultimate tensile strength (UTS) is typically $10\%$ below the in-plane compressive plateau stress $\sigma_{\rm Y}$ for both values of $\overline\rho$, as seen from \cref{fig:exp_tension_compression}. 

The Young's modulus $E$ of each foam is measured from the slope of the unloading curve of the in-plane compressive response at a small value of plastic strain on the order of $1\%$. Power-law fits to the measured mean values of $E$ and $\sigma_{\rm Y}$ provide scaling laws for $E$ and $\sigma_{\rm Y}$ in terms of the foam relative density $\overline\rho$:
\begin{equation}
    \dfrac{E}{E_{\rm S}} =  \overline\rho^{2.14} \quad \textrm{and} \quad \dfrac{\sigma_{\rm Y}}{\sigma_{\rm YS}} = 0.7\; \overline\rho^{1.71}
\label{eqn:scaling_e_sy}
\end{equation}
where $E_{\rm S}$ and $\sigma_{\rm YS}$ are the Young's modulus and yield strength, respectively, of the fully dense aluminium 6101 alloy. Here, $E_{\rm S}$ equals \SI{70}{GPa} and $\sigma_{\rm YS}$ equals \SI{200}{MPa}, as taken from \citet{Ashby2000}. The exponents on $\overline\rho$ in \cref{eqn:scaling_e_sy} are broadly consistent with bending of the cell walls of the foam: simple analytical models~\citep{Gibson1997} suggest $E \propto \overline\rho^2$ and $\sigma_{\rm Y} \propto \overline\rho^{3/2}$. 

The plastic \poisson ratio $\nu^{\rm P}$ of the foam was measured from the uniaxial compression tests using DIC software. The lateral strain in each specimen was measured at 5 equally spaced transverse sections of the specimen at a compressive strain of $10\%$. The mean value of $\nu^{\rm P}$ from 4 nominally identical specimens for each value of $\overline\rho$ was found to be $\nu^{\rm P}=0.17$ for $\overline\rho=6.6\%$ and $\nu^{\rm P}=0.25$ for $\overline\rho=9.6\%$.

\section{Fracture tests}
\subsection{Crack growth resistance curves}
The response of the deep notched bend specimen is shown in \cref{fig:expt_senb} in terms of the load $P$ and crack extension $\Delta{a}_{\rm PD}$ from DCPD as a function of cross-head displacement $v$. Data are shown for 4 repeat tests for $\overline\rho=6.6\%$ and 9.6\%, with the solid line corresponding to the mean response from the 4 tests and the shaded region representing the scatter in data. In all cases, crack growth (as inferred from the potential drop method) initiates prior to peak load. The accuracy of the potential drop method was verified by additionally measuring the crack extension by the unloading compliance method as described in \citet{e1820}. Acceptable agreement was found between the unloading compliance and potential drop measurements.

The \j$_{\rm R}$ versus $\Delta{a}_{\rm PD}$ crack growth resistance curves are obtained from the $P$ versus $v$ and $\Delta{a}_{\rm PD}$ versus $v$ curves following the steps outlined in \citet{e1820}; these responses are shown in  \cref{fig:exp_frac_ja} for both values of $\overline\rho$. Significant $R-$curve behaviour is observed for both values of relative density. The extent of the plastic zone ahead of the crack tip was measured using surface strain mapping by the DIC software.  Contours of strain $\varepsilon_{yy}$ are shown in \cref{fig:exp_frac_pzone} at peak load for one specimen of $\overline\rho=6.6\%$; it reveals the existence of a plastic zone on the order of the crack length at the onset of crack growth. The $R-$curve of the foam is attributed to two distinct zones of energy dissipation: (i) the bulk plastic zone emanating from the crack tip (containing a distribution of broken struts), and (ii) the crack bridging zone behind the advancing crack tip.

The value of toughness at crack initiation $J_{\rm IC}$ is determined from the $J-R$ resistance curves as follows. A crack blunting line, $J_{R} = 2 \sigma_{\rm Y} \Delta{a}$, is drawn as shown in \cref{fig:exp_frac_ja}, and an offset line is drawn parallel to the blunting line, intersecting the abscissa at $0.2$ mm. The value of the initiation toughness $J_{\rm IC}$ is given by the intersection of the $J-R$ curve with the $0.2$mm offset line. Following this procedure, we obtain $J_{\rm IC}=$\SI{0.5}{kNm^{-1}} for $\overline\rho=6.6\%$ and $J_{\rm IC}=$\SI{0.85}{kNm^{-1}} for $\overline\rho=9.6\%$. 

%------------------------------------------------------------------------------------------------------
% DAMAGE VISUALISATION USING CT
%------------------------------------------------------------------------------------------------------
\subsection{Extent of the process zone due to strut failure} \label{sec:bridging}
A distribution of failed struts forms a fracture process zone at the crack tip. The location of all failed struts at a given value of crack extension was obtained by X-ray computed tomography, as described previously in \cref{sec:damage-ct}. Consider one specimen of $\overline\rho=6.6\%$ in detail.  The distribution of failed struts projected over all $z$ in the $x-y$ plane, and projected over all $y$ in the $x-z$ plane is shown in \cref{fig:ERG_RelDen0066_Da2_failedStrutsCombinedZY_0066} for $\Delta{a_{\rm PD}} = \SI{2}{mm}$ and in  \cref{fig:ERG_RelDen0066_Da10_failedStrutsCombinedZY_0066_1} for $\Delta{a_{\rm PD}}=\SI{10}{mm}$. We emphasise that the markers in the projected $x-y$ plane indicate the $(x,y)$ location of the mid-point of each failed strut over all values of $z$, that is $0 \leq z \leq B$. Likewise, the markers in the projected $x-z$ plane indicate the $(x,z)$ location of the mid-point of each failed strut over for all values of $y$ such that $-L/2 \leq y \leq L/2$. {\rewrite{The damage parameter $f$ as a function of location $x$ at any value of $\Delta{a}_{\rm PD}$ is obtained as follows. We first identify the number of failed struts $n_{\rm f}$ within each control volume of $\Delta{W}\,L\,B$ (where $\Delta{W}=\SI{1.5}{mm}$, $L=\SI{10}{mm}$, and $B=\SI{26.4}{mm}$) corresponding to a traction-free fracture surface. The average value of $n_{\rm f}(x)$ is $22$, see for example \cref{fig:ERG_RelDen0066_Da10_failedStrutsCombinedZY_0066_1}, for control volumes at $x<\SI{6}{mm}$. Second, we count the number of failed struts $n(x)$ within each control volume centred at $x$. Then, the fraction of failed struts $f(x)$ is defined as $f = n/n_{\rm f}$.}} The limit $f=1$ is a somewhat severe requirement due to scatter in material ductility, and scatter in the value of $n_{\rm f}(x)$ in the definition of a traction-free crack, and so we arbitrarily assume that $f=0.9$ corresponds to a traction-free crack extension (i.e. no bridging ligaments) whereas $f=0$ corresponds to a region where no struts have failed. The distribution of failed struts ahead of the initial crack tip is shown in   \cref{fig:ERG_RelDen0066_Da10_failedStrutsCombinedZY_0066_2} for the choice $\Delta{a_{\rm PD}} = \SI{10}{mm}$. The region over which $0 \leq f \leq 0.9$ can be interpreted as a fracture process zone (FPZ), or equivalently a crack bridging zone (depending upon the assumed location of the crack tip). For definiteness, we shall assume that the physical crack tip exists at the transition point from $f \equiv 0$ to $f>0$. Define the maximum extent of the FPZ, $\Delta a_\mathrm{D}$, as the distance along the $x-$direction from the initial crack tip to the nearest location of $f=0$. The traction-free crack extension $\Delta a_\mathrm{f}$ is the length over which $f \geq 0.9$ ahead of the initial crack tip. We note from  \cref{fig:ERG_RelDen0066_Da10_failedStrutsCombinedZY_0066_2} that an inferred crack extension of $\Delta{a_{\rm PD}}=\SI{10}{mm}$ corresponds to traction-free extension of $\Delta a_\mathrm{f}=\SI{8.4}{mm}$ and a FPZ of length $\Delta{a_{\rm D}}=\SI{18.3}{mm}$ ahead of the current location of the traction-free crack tip. %In contrast, when $\Delta{a_{\rm PD}} = \SI{2}{mm}$, the bridging zone is of extent $\Delta{a_{\rm D}}=\SI{7.4}{mm}$ and the traction-free crack extension $\Delta{a_{\rm f}}$ does not exist. 

The extent of damage $f$ is plotted as a function of location $x$ ahead of the initial crack tip in \cref{fig:ERG_RelDen0066_2_failedStruts_stat} for selected values of $\Delta a_\mathrm{PD}$ in the range $\SI{1}{mm}$ to $\SI{10}{mm}$. The relation between $\Delta a_\mathrm{f}$, $\Delta a_\mathrm{D}$, and $\Delta a_\mathrm{PD}$ is given in 
 \cref{fig:ERG_RelDen0066_FreeAndDamageCrackTip}. Note that, when $\Delta{a_{\rm PD}} = \SI{2}{mm}$, the FPZ is of size $\Delta{a_{\rm D}}=\SI{7.4}{mm}$ and the traction-free crack extension $\Delta{a_{\rm f}}$ vanishes. In general, the traction-free crack tip lags behind the inferred crack tip from PD measurements by about $\SI{4}{mm}$.  Further, the extent of the bridging zone $\Delta a_\mathrm{D}$ increases steeply with crack extension: $\Delta a_\mathrm{D}=\SI{3.4}{mm}$ at $\Delta a_\mathrm{PD}=\SI{1}{mm}$ and $\Delta a_\mathrm{D}=\SI{18.3}{mm}$ at $\Delta a_\mathrm{PD}=\SI{10}{mm}$. 
It is instructive to compare the relative extent of crack tip plastic zone $r_{\rm P}$ and the FPZ $\Delta{a}_{\rm D}$ in order to assess whether a crack tip \j-field exists. As an approximation, assume that the foam yields when the von Mises measure of strain exceeds a value of $\sigma_{\rm Y}/E$, and define $r_{\rm P}$ as the maximum extent of plastic zone from the crack tip. The von Mises strain $\varepsilon_\mathrm{e}$ was calculated from the measured values of minor principal strain $\varepsilon_1$ and the major principal strain $\varepsilon_2$ on the $z=0$ plane as
\begin{equation}
    \varepsilon_\mathrm{e}^2={\frac{2}{9}\left[ (\varepsilon_1 - \varepsilon_2)^2 + \varepsilon_1^2 + \varepsilon_2^2 \right]} 
    \label{eqn:rp_strain}
\end{equation}
assuming plane strain, $\varepsilon_3 \equiv 0$. The values of $\varepsilon_1$ and $\varepsilon_2$ were obtained from the DIC software based on facets of size on the order of the foam cell size. The extent of plastic zone $r_{\rm P}$ is plotted in \cref{fig:ERG_RelDen0066_FreeAndDamageCrackTip} as a function of crack extension for $\overline\rho=6.6\%$. Plots similar to \cref{fig:ERG_RelDen0066_2_failedStruts_stat,fig:ERG_RelDen0066_FreeAndDamageCrackTip} have also been generated for the case of $\overline\rho=9.6\%$, see \cref{fig:ERG_RelDen0096_2_failedStruts_stat,fig:ERG_RelDen0096_FreeAndDamageCrackTip}.

We conclude from \cref{fig:ERG_RelDen0066_FreeAndDamageCrackTip,fig:ERG_RelDen0096_FreeAndDamageCrackTip} that the FPZ size $\Delta{a}_{\rm D}$ in the Al alloy foams is on the order of the plastic zone size at the onset of traction-free crack extension, and for subsequent crack growth. Consequently, a crack tip $J-$field does not exist, and the measured value of $J_{\rm IC}$ following the ASTM $J-$integral test procedure cannot be treated as a material property despite the specimen size meeting the criteria of the ASTM standard. We emphasise that the ASTM procedure is based on the assumption that the FPZ is much smaller than the plastic zone size. This criterion is obeyed in fully dense metals but is violated for the metal foam under current consideration. Note that the fracture toughness $K_{\rm IC}$ remains a valid material parameter for the foam provided the specimen size is sufficiently large for an outer $K-$field to exist. Recall that the ASTM guideline suggests the following dimensions for a valid $K_{\rm IC}$ test on a single edge notch bend geometry:
\begin{equation}
    \left(a_0, W-a_0, B\right)> 2.5 \left( \dfrac{K_{\rm IC}}{\sigma_{\rm Y}}  \right)^2
\label{eqn:kic_size}
\end{equation}
This size criterion remains meaningful, but we cannot estimate the value of $K_{\rm IC}$ from the value of $J_{\rm IC}$ from a \j$-$test. {\rewrite{We emphasise that this case of metals foams is different from situations such as those discussed by \citet{Pineau1981} where $J$ is an insufficient parameter to characterise fracture in fully dense metals. In those cases a $J-$field exists but fracture is dependent on both the $J-$field and the so-called non-singular $Q$ term, introduced by O'Dowd and Shih \citep{Odowd1991, Odowd1992}.}}

\subsection{Conditions for the existence of a crack tip $J-$field: fully dense alloys versus micro-architected materials}
Recall that the criterion for the existence of a crack tip $J-$field in a fully dense alloy is given by \cref{eqn:sizecriterion}, consistent with \cref{fig:sketch_ssb_lsb}(a). Additional length scales arise in metallic foams (and in lattice materials in general): the cell size $\ell$ and strut thickness $t$. The criterion for the existence of a crack tip $J-$field may thus differ from that stated in \cref{eqn:sizecriterion}. A $J-$field can only exist if the plastic zone spans a minimum number $n_1$, of cells, such that $r_{\rm P} > n_1 \ell$ (where the precise value of $n_1$ requires future study). Further, the FPZ must be contained within the zone of $J-$dominance. This criterion is achievable when fracture occurs in a sequential fashion strut-by-strut at the crack tip. In contrast, when struts fail stochastically in a zone of size comparable to that of the plastic zone, an annular zone of $J-$dominance does not surround the FPZ, recall \cref{fig:sketch_ssb_lsb}(b). In this case, no near-tip $J-$field exists and $J$ cannot be used as a fracture parameter. This is the case for the current metallic foam under consideration. 

%------------------------------------------------------------------------------------------------------
% SIMULATIONS
%----------------------------------------------------------------------------------------------------
\section{Fracture model} \label{sec:numerical_approach}
Can a cohesive zone model be used to predict the fracture response of the foam of the present study? To address this, the cohesive zone method of \citet{Tvergaard1992} is used to model crack advance in the deep-notched bend (SENB) specimen. This method allows for crack bridging in the presence of bulk plasticity as observed in the metal foam without assuming \textit{a priori} that the bridging zone is much smaller than the plastic zone. Static finite element (FE) calculations were performed with ABAQUS/Standard v6.14 to aid interpretation of the experimental observations such as the $R-$curve and the bridging zone. The foam is modelled as an isotropic, compressible elastic{-}plastic strain-hardening solid, based on the compressible elastic{-}plastic constitutive model of \citet{Deshpande2000}.  The objectives of the numerical study are: (i) to deduce the cohesive parameters such as the peak traction \shat and toughness \gc based on the measured load versus displacement response, and (ii) to determine whether the cohesive zone model can predict the extent of crack growth. 

\subsection{FE model}
The FE mesh assumes that the notch in the SENB specimen has a semi-circular tip\footnote{A series of additional FE simulations were performed with a sharp crack tip. It was found that the notch acuity has a negligible effect upon the load versus displacement response.} of diameter $d = \ell$, where $\ell=\SI{2}{mm}$ is the average cell edge length of the foam. Introduce the co-ordinate system $(x,y)$ as shown in \cref{fig:fe_senb_geometry}. A cohesive zone is placed ahead of notch on $a_0 \leq x \leq W$ comprising four-noded cohesive elements (type COH2D4 in ABAQUS) of zero thickness. The FE mesh for the foam comprises linear quadrilateral elements of plane strain (type CPE4R). All rollers are modelled as rigid surfaces, and frictionless contact is assumed between the rollers and foam. A symmetric half model is employed in the FE study with the support roller fixed in all directions and the loading roller subjected to a constant downward velocity in the $z-$direction. The velocity of loading is chosen to be sufficiently small for inertial effects to be negligible; the response obtained from the explicit FE simulation is thus quasi-static. Salient features of the assumed material models for the foam and the cohesive zone are outlined below.

\subsection{Material model for the foam}
The initial modulus of the foam (for $\overline\rho=6.6\%$) is taken to be $E=\SI{175}{MPa}$, based on the measured mean value of the unloading modulus during uniaxial compression test.  {\rewrite{An elastic Poisson's ratio of $0.3$ is assumed, based on \cite{Gibson1997}}}. The post-yield behaviour of the foam is modelled using the ABAQUS crushable foam model with isotropic hardening, based on the constitutive model of \citet{Deshpande2000}, as follows. 
The yield surface is assumed to be elliptical, with the centre of the ellipse located at the origin of the mean stress versus von Mises effective stress plane. It evolves in a geometrically self-similar manner, and is of the form
\begin{equation}
    \phi = \hat\sigma - \sigma_{\rm Y} = 0
    \label{eqn:yieldsurface}
\end{equation}
where $\sigma_{\rm Y}$ is the uniaxial yield strength of the foam (assumed to be identical in tension and compression), and the effective stress $\hat\sigma$ is defined as
\begin{equation}
    \hat\sigma^2 = \dfrac{1}{1 + \left( \dfrac{\alpha}{3} \right)^2} \left(  \sigma_{\rm e}^2 + \alpha^2 \; \sigma_{\rm m}^2  \right) 
\label{eqn:ellipse}
\end{equation}
Here, $\sigma_{\rm e}$ is the von Mises effective stress, $\sigma_{\rm m}$  is the mean stress,   and $\alpha$ is the shape factor of the yield ellipse. Associated plastic flow rule is assumed. Consequently, $\alpha$ is related to the plastic Poisson's ratio \nup of the foam  according to
\begin{equation}
\alpha^2 =   \dfrac{9}{2} \left({ \dfrac{1-2\nu^{\rm P}}{1+\nu^{\rm P}} } \right)
\label{eqn:alpha-mu}
\end{equation}
During plastic flow, the yield surface grows in a geometrically self-similar manner with strain, in accordance with the specified hardening response and the fixed value of shape factor $\alpha$ of the yield surface. 

The initial yield strength $\sigma_{\rm Y}=\SI{1.42}{MPa}$ of the foam  (for $\overline\rho=6.6\%$) is based on the measured mean value from 4 uniaxial compression tests, recall \cref{fig:exp_tension_compression}. The assumed (idealised) true stress $\sigma$ versus true strain $\varepsilon$ response for the foam is shown in \cref{fig:fe_material}; it is derived from the mean of the measured nominal compressive stress $\sigma_{\rm n}$ versus nominal compressive strain $\varepsilon_{\rm n}$ responses of \cref{fig:exp_tension_compression} using the following relation to account for plastic compressibilty:
\begin{equation}
    \sigma= \sigma_{\rm n}  \left[  \dfrac{1+\varepsilon_{\rm n}} {1+ \left( 1- 2 \nu^{\rm P} \right)\varepsilon_{\rm n} }   \right]
\end{equation}
A value of \nup$=0.17$ is assumed for the foam of $\overline\rho=6.6\%$ based on the measured mean value of \nup from 4 tests, recall \cref{sec:material_charac}. The shape factor $\alpha$ of the yield surface follows from \cref{eqn:alpha-mu} as $\alpha=1.58$.

\subsection{Cohesive zone model}
Crack advance from the tip of the pre-notch is modelled via the tensile traction versus opening displacement relation. It is assumed that crack growth occurs on the symmetry plane $y=0$. Consequently, it suffices to specify a relation for the normal traction $T$ and crack opening $\delta$. We adopt the Xu–Needleman interfacial law \cite{Xu1994} for each cohesive element in both tension and compression; it has the form
\begin{equation}
    \dfrac{T}{\hat\sigma} = \dfrac{\delta}{\delta_{\rm n}} \exp\left(  1- \dfrac{\delta}{\delta_{\rm n}} \right)
    \label{eqn:czlaw}
\end{equation}
where $\hat\sigma$ is the peak crack opening traction that occurs at an opening $\delta=\delta_{\rm n}$, as shown in \cref{fig:coh-law}. The fracture energy (or toughness) of the cohesive zone is the area under the $T$ versus $\delta$ curve: $\Gamma_0 = \int_{0}^{\infty} T \; d\delta =  e \hat\sigma \delta_{\rm n}$. The crack opening displacement  $\delta$ is related directly to the material displacement along the centre-line of the specimen such that $\delta = 2u_{y}\left( 0^+, x \right)= - 2u_{y}\left( 0^-, x \right)$. 

The precise values of the cohesive properties $(\hat\sigma, \Gamma_0)$ for the foam are not known \textit{a priori}; instead, we deduce values for $(\hat\sigma, \Gamma_0)$ based on a goodness-of-fit between the predicted and measured load versus displacement response during crack growth. To achieve this, a series of FE simulations were performed for selected values of $\hat\sigma$ from $\SI{1.42}{MPa}$ (equal to the initial yield strength of the solid $\sigma_{\rm Y}$) to $\SI{2}{MPa}$, and $\Gamma_0$ in the range of $\SI{0.1}{kNm^{-1}}$ to $\SI{1}{kNm^{-1}}$.  Recall that the characteristic cohesive length $\ell_{\rm c}$ is defined by 
\begin{equation}
    \ell_{\rm c} =\dfrac{\pi}{8} \dfrac{E\Gamma_0}{\hat\sigma^2} 
    \label{eqn:dugdale}
\end{equation}
In order to ensure adequate mesh resolution during all stages of crack growth, $0 \leq \Delta{a} \leq W-a_0$, a uniform FE mesh of element size $\ell_{\rm e}$ (for both the solid and the cohesive zone) is constructed across the ligament such that $\ell_{\rm e} \leq 0.05\; \ell_{\rm c}$, see \cref{fig:fe_senb_geometry}. 

\section{Predicted fracture response} \label{sec:predictions}
\subsection{Validation of the foam constitutive model}
In order to verify the accuracy of the foam constitutive model, an FE simulation is performed for the case of 3-point bending of the foam specimen absent a pre-notch (and without a cohesive zone). The relevant dimensions of this specimen are: $S=\SI{200}{mm}$, $W=\SI{30}{mm}$, and plane strain thickness $B=\SI{26.4}{mm}$. 

The measured mean curve of the load versus displacement response is shown in \cref{fig:fe_bend}; the load $P$ is normalised by the plastic collapse load, $\sigma_{\rm Y}BW^2/S$,  and the roller displacement $v$ is normalised by the span $S$. The peak load of the bend specimen is dictated by  plastic collapse of the ligament of the specimen. Subsequently, tensile failure of the cell walls occurs at the outermost layer, and a crack propagates through the specimen, leading to a softening $P$ versus $v$ response. The FE prediction of the bend response is included in \cref{fig:fe_bend}. Good agreement is noted between the FE prediction and the experiment until plastic collapse of the section at the mid-length ($y=0$) occurs at $PS/\sigma_{\rm Y}BW^2 = 1$. Failure of the foam was not included in the FE model; the continued mild hardening in $P$ versus $v$ response for $v/S>0.03$ is a consequence of the assumed strain-hardening response of the foam. 

\subsection{Response of a deep notch specimen under 3-point bending: limiting cases}
Consider the following 3 limiting cases of cohesive zone properties: (i) $\hat\sigma=\Gamma_0=\infty$ such that softening and crack extension do not occur, (ii) a rigid, ideally plastic cohesive zone with $ \hat\sigma=\sigma_{\rm Y}$ and $\Gamma_0=\infty$, and (iii) $\hat\sigma=\sigma_{\rm Y}$ and $\Gamma_0 =\SI{0.1}{kNm^{-1}}$ to simulate brittle fracture. (Note that the choice $\Gamma_0 = 0$ would require $\ell_{\rm e} = 0$ in order to adequately resolve the crack tip field within the cohesive elements.) Denote the initial ligament length by $b_0$ such that $b_0=W-a_0$. The normalised load $PS/\sigma_{\rm Y}Bb_0^2$ versus normalised displacement $v/S$ for these limiting cases are plotted in \cref{fig:limitingcases} and they demonstrate the bounds of the $P$ versus $v$ response that can be obtained from the cohesive zone model. The measured mean response lies within these bounds. The responses of cases (i) and (ii) are nearly identical due to the bulk plasticity at the notch tip and beneath the roller. The mild increase in load beyond the plastic collapse load $\left(PS/\sigma_{\rm Y}Bb_0^2 = 1\right)$ is due to the strain-hardening characteristic of the foam. 

\subsection{Extraction of the cohesive parameters}
A series of FE simulations were performed using selected combinations of $(\hat\sigma, \Gamma_0)$, with $\hat\sigma$ between $\SI{1.42}{MPa}$ and $\SI{2}{MPa}$, and $\Gamma_0$ between $\SI{0.1}{kNm^{-1}}$ and $\SI{1}{kNm^{-1}}$. Each FE simulation, obtained for a given combination of $(\hat\sigma, \Gamma_0)$, gives rise to the following quantities which can be compared with the experimental observations:
\begin{itemize}[leftmargin=2em]
\item [(i)] load $P$ versus roller (or cross-head) displacement $v$ response,
\item [(ii)] traction-free crack extension $\Delta{a}_{\rm f}$ versus $v$,
\item [(iii)] evolution of the crack tip opening displacement $\delta_{\rm T}$ and crack mouth opening displacement $\delta_{\rm M}$ with increasing $v$, 
\item [(iv)] FPZ size $\Delta{a}_{\rm D}$ (comprising cell wall failure and crack bridging), plastic zone  size $r_{\rm P}$, and  traction-free crack extension $\Delta{a}_{\rm f}$ ahead of the pre-notch tip, each versus $v$.
\end{itemize}
The crack tip opening displacement $\delta_{\rm T}$ and crack mouth opening displacement $\delta_{\rm M}$ are defined in \cref{fig:fe_senb_geometry}:  $\delta_{\rm T}$ is the change in distance between points T and T' placed on the diametric ends of the semi-circular notch tip, and $\delta_{\rm M}$ is the change in  distance between points M and M' at the notch mouth. The precise locations of T, T', M, and M' are identified (within $\pm \SI{0.1}{mm}$) and tracked during the experiment using DIC software. Traction-free crack extension is assumed to occur in the FE simulation when the traction $T$ at an integration point within the cohesive element drops to $0.01\hat\sigma$. The traction-free crack extension $\Delta{a}_{\rm f}$ in the experiments is determined from the 3D reconstruction of the fracture process zone based on a set of XCT scans, recall \cref{fig:ERG_RelDen0066_2_failedStruts_stat}. 

In order to extract the cohesive parameters $(\hat\sigma, \Gamma_0)$ associated with the fracture process, we define a goodness-of-fit measure, $\chi$, for the $P$ versus $v$ response as follows:
\begin{equation}
    \chi = \dfrac{\int_0^{v_{\rm f}} | P_{\rm exp}\left(v \right) - P_{\rm FE}\left(v \right) | \; dv}{\int_0^{v_{\rm f}} P_{\rm exp} \left(v \right) \; dv}
\end{equation}
Here, $P_{\rm exp}\left(v \right)$ is the load versus displacement response from the experiment; a mean response from 4 tests is assumed for the calculation of $\chi$. $P_{\rm FE}\left(v \right)$ is the predicted load versus displacement $v$ response and  $v_{\rm f}$ is the maximum value of cross-head displacement $v$ in the experiment. A value of $\chi=1$ indicates perfect agreement between predicted and measured responses. Contours of $\chi$, as obtained from a set of $64$ FE simulations, are plotted in \cref{fig:contours_chi} with axes of cohesive strength $\hat\sigma$ and toughness $\Gamma_0$. 

Three distinct local maxima of best fit ($\chi \geq 0.9$) emerge in the map of \cref{fig:contours_chi}. We direct our attention towards the optimal point within each of these regions: point A with $(\chi, \hat\sigma, \Gamma_0) = (0.91, \SI{1.47}{MPa}, \SI{0.96}{kNm^{-1}})$, point B with $(\chi, \hat\sigma, \Gamma_0) = (0.94, \SI{1.63}{MPa}, \SI{0.45}{kNm^{-1}})$, and point C with $(\chi, \hat\sigma, \Gamma_0) = (0.97, \SI{2.00}{MPa}, \SI{0.10}{kNm^{-1}})$. The corresponding cohesive laws for the 3 cases are shown in \cref{fig:bestfit_cohlaw}. The predicted load versus displacement response for cases A, B, and C are compared in \cref{fig:bestfit_load} with the measured response. Note that the Dugdale plastic zone length as given by \cref{eqn:dugdale} is of magnitude $\ell_{\rm c} = \SI{30.5}{mm}$, $\SI{11.7}{mm}$ and $\SI{1.7}{mm}$ for cases A, B and C respectively. 

The crack tip opening displacement $\delta_{\rm T}$ and crack mouth opening displacement $\delta_{\rm M}$ for the three best fitting cases A, B, and C are almost indistinguishable, and they agree well with the corresponding measured data from DIC, see \cref{fig:bestfit_cmod}. Thus, it is not possible to distinguish the best choice of $(\hat\sigma, \Gamma_0)$ values on the basis of $P$ versus $v$ response alone. In order to gain further insight into the best choice of A, B, or C, it is necessary to explore the accuracy of the predictions with additionally available experimental data, as follows.

Consider the cohesive law of \cref{fig:coh-law}. We assume that damage in the cohesive element initiates when the traction $T$ attains the peak value of $\hat\sigma$; the damage parameter $f$ at this instant is equal to zero. With increasing opening displacement $\delta$, $f$ increases until it becomes unity when $T$ drops to $0.01\hat\sigma$, marking the onset of traction-free crack extension $\Delta a_\mathrm{f}$. The region over which $0 \leq f < 0.9$ in the cohesive elements ahead of the pre-crack can be interpreted as the fracture process zone (FPZ), recall \cref{sec:bridging}. Consequently, the maximum extent of the FPZ, $\Delta a_\mathrm{D}$, corresponds to the distance along the ligament from the initial crack tip to the location of $f=0$. Likewise, the traction-free crack extension $\Delta a_\mathrm{f}$ is the distance along the ligament from the initial crack tip over which $f$ attains or exceeds $0.9$. 
The extent of damage zone $\Delta{a}_{\rm D}$ and the traction-free crack extension $\Delta{a}_{\rm f}$ from the FE simulations are shown in \cref{fig:bestfit_dadamage,fig:bestfit_da}, respectively, as a function of roller displacement $v$, for the 3 best fitting cohesive laws. The measured values of $\Delta{a}_{\rm D}$  and $\Delta{a}_{\rm f}$, as taken from \cref{fig:exp_frac_av} and \cref{fig:ERG_RelDen0066_FreeAndDamageCrackTip}, are included in \cref{fig:bestfit_dadamage,fig:bestfit_da}, respectively.  The ordinate in each case is normalised by the initial ligament length, $b_0=W-a_0$. We note from \cref{fig:bestfit_dadamage,fig:bestfit_da} that case A best predicts the evolution of damage zone $\Delta{a}_{\rm D}$ but case B provides best agreement with the observed traction-free crack extension $\Delta{a}_{\rm f}$. The cross-plot of $\Delta{a}_{\rm f}$ and $\Delta{a}_{\rm D}$ in \cref{fig:bestfit_dafreedamage} further demonstrates the large variation in response for cases A to C. 

It is instructive to compare the extent of the plastic zone $r_{\rm P}$ at the tip of the pre-crack to the size of the FPZ.  Predictions of $r_{\rm P}$ based on \cref{eqn:rp_strain} are shown in \cref{fig:bestfit_pzdadamage} for the 3 best fitting cohesive laws along with the DIC measurement of $r_{\rm P}$. The FPZ size $\Delta{a}_{\rm D}$ is generally smaller than  the plastic zone size $r_{\rm P}$, with the observed response sandwiched between the predictions for cases A and B.  Further, we find from \cref{fig:bestfit_dafreedamage,fig:bestfit_pzdadamage} that the initiation of crack growth $\left(\Delta{a}_{\rm f}=0^+\right)$ is accompanied by a large FPZ in all 3 predictions (as well as in the experiment):  $\Delta{a}_{\rm D} / r_{\rm P} = 0.92$ for case A, $\Delta{a}_{\rm D} / r_{\rm P} = 0.36$ for case B, and $\Delta{a}_{\rm D} / r_{\rm P} = 0.1$ for case C, at $\Delta{a}_{\rm f}=0^+$. These points are marked by the symbol $\textsf{O}$ in \cref{fig:bestfit_pzdadamage} for clarity.

We conclude from \cref{fig:fe_bestfit} that no unique pair of 
$(\hat\sigma, \Gamma_0)$ examined here can simultaneously capture all the  experimental observations: load, crack extension, and the development of the damage/bridging zone. The predictions of crack growth are sensitive to the choice of $\hat\sigma/\sigma_{\rm Y}$: a value of $\hat\sigma$ less than $\sigma_{\rm Y}$ implies that the fracture process zone is embedded within an elastic solid, thereby leading to a flat $R-$curve. In contrast, a value of $\hat\sigma$ greater than $\sigma_{\rm Y}$ will give an increasing $R-$curve but the low strain hardening characteristic of the foam leads to an extreme sensitivity of the predicted response to the precise choice of $\hat\sigma/\sigma_{\rm Y}$. Additionally, we find that the values of $\hat\sigma/\sigma_{\rm Y}$ that give best alignment between predicted and measured load versus displacement (as well as crack mouth and crack tip opening displacements) corresponds to tensile strains on the order of $40\%-50\%$ at the notch-tip; this is unrealistic for metal foams since their tensile ductility is on the order of a few percent ($2\%-5\%$, depending upon the relative density). 

%------------------------------------------------------------------------------------------------------
% DENT TEST: EXPERIMENT AND PREDICTION
%------------------------------------------------------------------------------------------------------
\section{An attempt to measure directly the cohesive zone law}
We proceed to explore whether the  mode I cohesive law can be measured directly from the response of a deep-notch tensile specimen. In particular, the relationship between the cohesive zone law and the average traction versus additional axial displacement of a deeply notched specimen, is now determined. Here, we interpret the additional displacement as the additional elongation associated with the presence of the edge cracks and plasticity in the net section.

Double edge notch specimens of length $2L=\SI{100}{mm}$, width $2W=\SI{50}{mm}$, and thickness $B=\SI{26.4}{mm}$ were  machined from a flat panel of $\overline\rho=6.6\%$ foam. Notches of length $a$ were machined  using a fine blade of thickness $\SI{300}{\micro m}$, on both sides of the specimen, see \cref{fig:dent_geometry}. Two notch sizes were considered: $a=\SI{15}{mm}$ and $a=\SI{20}{mm}$. The extra displacement $\Delta{u}$ associated with plasticity and distributed cracking of the net section between the notches was measured using DIC by tracking two facet points spaced $\SI{10}{mm}$ apart and symmetrically about the mid-plane as shown in \cref{fig:dent_geometry}.

The load versus displacement response of the two deep notched specimens are shown in \cref{fig:dent_load}. The net section stress $\sigma^\infty$ in both cases is normalised by the un-notched compressive yield strength $\sigma_{\rm Y}$ of the foam, and is plotted against the gauge displacement $\Delta{u}$. The peak value of $\sigma^\infty$ for both the notch geometries slightly exceeds the yield strength of the foam, consistent with the notch strengthening behaviour observed in open-cell metallic foams due to size effects, see for example, \citet{Andrews2001} and \citet{Combaz2011}. 

Contours of the displacement $u_{y}$ in the loading direction at peak load, as obtained from the DIC, are shown in \cref{fig:dent_u1}(a) and (b), for $a=\SI{15}{mm}$ and $a=\SI{20}{mm}$, respectively. Note that there is no significant displacement jump within a thin ligament between the notch tips. Consequently, the measured  $\sigma^\infty$ versus $\Delta{u}$ response is not a direct measure of the crack bridging law. The DIC contours show a displacement jump $\Delta{u_{y}}$ of $\SI{0.4}{mm}$ across the gauge length  of $\SI{10}{mm}$ at peak load: the observed failure strain of $4\%$ from the DIC is consistent with the tensile ductility of the foam.

\subsection{An independent test that makes use of the calibrated cohesive law}
FE simulations were performed on the two double edge notch specimens ($a/W=0.6$ and $a/W=0.8$)  by using the best calibration, case B, from the bend tests for the cohesive zone parameters. The $T$ versus $\delta$ curve for case B is included in \cref{fig:dent_u1}(b); it differs from the measured $\sigma^\infty$ versus $\Delta{u}$ response in the softening portion of the curve.  Predictions of the load versus displacement response are shown in \cref{fig:dent_load} along with contours of displacement $u_{y}$ in \cref{fig:dent_u1}(a) and (b). We find that the cohesive zone model gives an acceptable agreement of the load versus displacement response for the two deep notch tensile specimens with their corresponding measured response. However, in a similar manner to the bend test, the FE simulation predicts unrealistically high notch root strains ($>20\%$) within a zone on the order of the cell size, see \cref{fig:dent_u1}(c) for contours of strain at peak load.

%------------------------------------------------------------------------------------------------------
% DISCUSSION/CONCLUSIONS
%------------------------------------------------------------------------------------------------------
\section{Concluding remarks}
The \j-test procedure, as outlined in the ASTM standard E1820, is valid only when the fracture process zone is much smaller than the plastic zone surrounding the crack tip. This is not the case for the aluminium alloy foams of the present study. A crack tip  \j-field does not exist for the foams studied here, at any stage of crack growth. This is traced to the fact that a large fracture process zone exists ahead of the crack tip, and extends to almost the plastic zone boundary. The measured value of $J_{\rm IC}$ is therefore not a material property.  The present experimental study does not consider specimens that are sufficiently large to satisfy the ASTM criterion for a remote $K-$field to exist, recall \cref{eqn:kic_size}. If tests were performed on specimens of sufficient size that an outer $K-$field exists, then this test could be used to measure $K_{\rm IC}$. Thus, $K_{\rm IC}$ remains a material property for the foam.

A cohesive zone model has the ability to capture the large scale bridging that occurs in metal foams, but it remains a challenge for the model to capture the main features of crack advance that are observed in the experiment in addition to the load versus displacement collapse response. In the present study, the cohesive zone model of \citet{Tvergaard1992}, along with a compressible plasticity model for the foam, was used to model crack growth in a deep notched bend specimen, and thereby used to extract the cohesive zone parameters: the cohesive strength and toughness. Three possible combinations of cohesive strength and toughness give an acceptable agreement with the measured load versus displacement response. However, the value of $\hat\sigma/\sigma_{\rm Y}$ that gives best agreement with the measured load versus displacement (as well as crack mouth and crack tip opening displacements) demands the existence of tensile strains on the order of $40\%-50\%$ at the notch-tip. This is unrealistic since the foam has a tensile ductility of only a few percent. None of the 3 choices of cohesive zone law was able to predict accurately the degree of crack extension (bridged crack and traction-free crack) as a function of remote displacement.

\section*{Acknowledgements}
The authors gratefully acknowledge financial support from the European Research Council (ERC) in the form of advanced grant, MULTILAT, GA669764. We are also grateful for enlightening discussions with Prof. John W. Hutchinson on the topic of cohesive zone modelling.

%------------------------------------------------------------------------------------------------------
% BIBLIOGRAPHY
%------------------------------------------------------------------------------------------------------
\bibliographystyle{elsarticle-num-names} 
\biboptions{sort&compress}
\bibliography{References}

%------------------------------------------------------------------------------------------------------
% FIGURES
%------------------------------------------------------------------------------------------------------
\input{Figures.tex}

%------------------------------------------------------------------------------------------------------
\end{document}

%% file: Preamble.tex
\usepackage{lineno}
\usepackage[a4paper, top=2.25cm, bottom=2cm, left=2.25cm, right=2.25cm]{geometry}
\modulolinenumbers[5]
\journal{Journal of Mechanics and Physics of Solids}
\usepackage{graphicx,ulem}
\usepackage{xspace} 
\xspaceaddexceptions{)}
\xspaceaddexceptions{]}
\usepackage[]{lmodern}
\usepackage{siunitx} 
\usepackage{tabularx}
\usepackage{floatrow}
\usepackage[labelfont={large}, labelformat=simple]{subfig}
\usepackage{color}
\usepackage{soul}
\usepackage{booktabs}
\usepackage{chngcntr}
\biboptions{numbers,sort&compress}
\usepackage{listings}
\usepackage{paralist}
\usepackage{epstopdf}
\usepackage{setspace}
\usepackage[export]{adjustbox}

\usepackage[scaled=1.05]{gentium} 
\newsavebox{\oldell} \savebox{\oldell}{\ensuremath{\ell}}
\let\temp\rmdefault \usepackage{kpfonts} \let\rmdefault\temp
\renewcommand*{\ell}{\usebox{\oldell}}
\usepackage[T1]{fontenc} 

\usepackage{fancyhdr}
\usepackage[colorlinks, plainpages=false]{hyperref}
\hypersetup{final} 
\usepackage{amsmath}
\usepackage[capitalise]{cleveref}
\let\subtef\subref
\renewcommand{\subref}[1]{\protect\subtef*{#1}}

\renewcommand{\bar}[1]{\mkern 1.25mu\overline{\mkern-1.25mu#1\mkern-1.25mu}\mkern 1.25mu}

\def\j{\textit{J}}

\def\shat{$\hat\sigma$\xspace }
\def\gc{$\Gamma_0$\xspace }
\def\nup{$\nu^{\rm P}$\xspace}
\def\poisson{Poisson\textquotesingle s }

\crefmultiformat{figure}%
{\edef\crefstripprefixinfo{#1}Figs.~#2#1#3}%
{ and~(#2\crefstripprefix{\crefstripprefixinfo}{#1}#3}%
{, #2\crefstripprefix{\crefstripprefixinfo}{#1}#3}%
{, and~#2\crefstripprefix{\crefstripprefixinfo}{#1}#3}

\usepackage{enumitem}  
\usepackage{titlesec}
\titlespacing{\subsection}{0pt}{\parskip}{-\parskip}
\titlespacing{\subsubsection}{0pt}{\parskip}{-\parskip}

\usepackage{pdfpages}
\usepackage{afterpage}

\makeatletter
\newcounter{savesection}
\newcounter{apdxsection}
\newcommand\unappendix{\par
  \setcounter{apdxsection}{\value{section}}%
  \setcounter{section}{\value{savesection}}%
  \setcounter{subsection}{0}%
  \gdef\thesection{\@arabic\c@section}}
\makeatother

\makeatletter
\providecommand{\doi}[1]{%
  \begingroup
    \let\bibinfo\@secondoftwo
    \urlstyle{rm}%
    \href{http://dx.doi.org/#1}{%
      \discretionary{}{}{}%
      \nolinkurl{#1}%
    }%
  \endgroup
}
\makeatother

\newcommand{\rewrite}[1]{\color{black}{ {#1}}}

\titleformat{\subsection}{\normalfont\bf}{\thesubsection}{1em}{}

%% file: Figures.tex
\newpage
\hypersetup{linkcolor = black}
\listoffigures

\newpage
\section*{Figures}
\counterwithout{figure}{section}
\setcounter{figure}{0}
\floatsetup[figure]{subcapbesideposition=top}
\setlength{\labelsep}{0cm}

\graphicspath{{/}}

%------------------------------------------------------------------------------------------------------
% SKETCH - J DOMNINANCE
%------------------------------------------------------------------------------------------------------
\begin{figure}[H]
\includegraphics[scale=1]{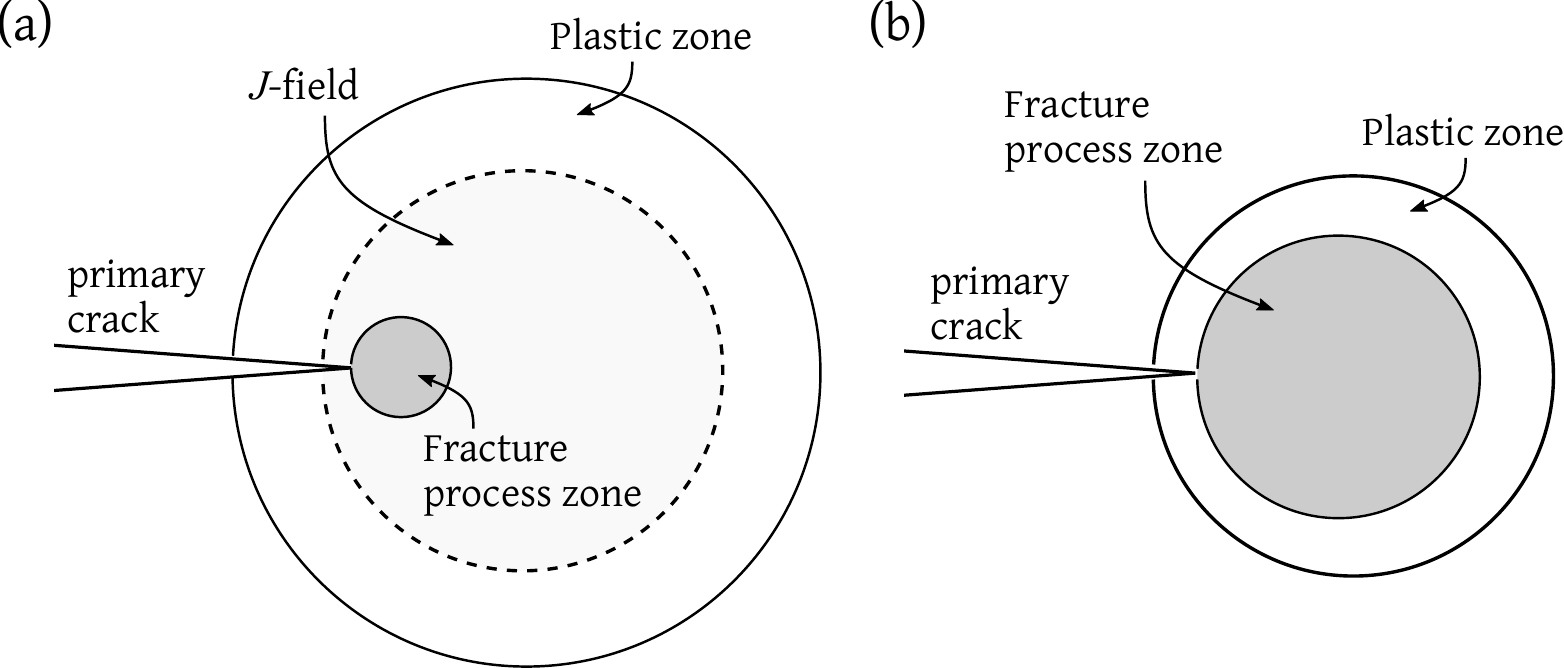}
\vspace{5mm}
\caption{\textbf{(a)} A fracture process zone (FPZ) embedded within a crack tip \j-field; \textbf{(b)} no crack tip \j-field is present due to the existence of a FPZ comparable in dimension to the plastic zone size.}
\label{fig:sketch_ssb_lsb}
\end{figure}  

%------------------------------------------------------------------------------------------------------
% CT SCANS OF FOAM MICROSTRUCTURE
%------------------------------------------------------------------------------------------------------
\vspace{10mm}
\begin{figure}[H]
  \setlength{\labelsep}{0.35cm}
  \sidesubfloat[]{\includegraphics[width=0.43\textwidth]{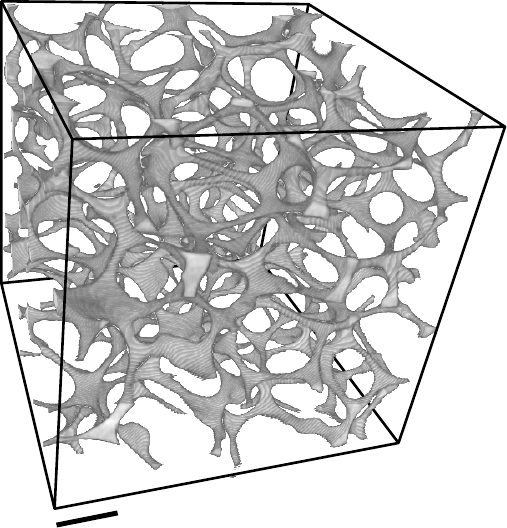}\label{fig:3D_RelDen0066}}\hfill
  \sidesubfloat[]{\includegraphics[width=0.43\textwidth]{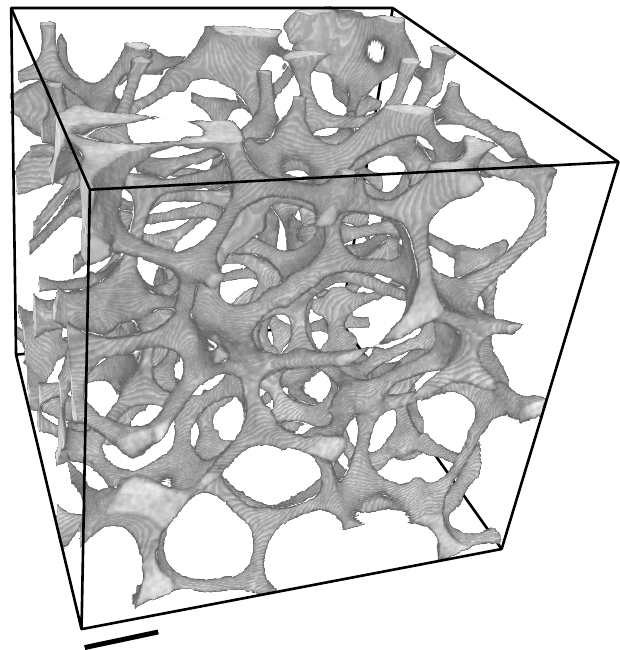}\label{fig:3D_RelDen0096}}
  \caption{X-ray CT images of aluminium alloy foam of
    \subref{fig:3D_RelDen0066} relative density $\bar{\rho}=6.6\%$ and \subref{fig:3D_RelDen0096}  $\bar{\rho}=9.6\%$. The scale bar is
    of length \SI{1}{mm}.}
  \label{fig:3D_RelDen}
\end{figure}

%------------------------------------------------------------------------------------------------------
% SPECIMEN GEOMETRIES: TENSION, COMPRESSION, BEND TOUGHNESS
%------------------------------------------------------------------------------------------------------
\newpage
\begin{figure}[H]
\includegraphics[scale=1]{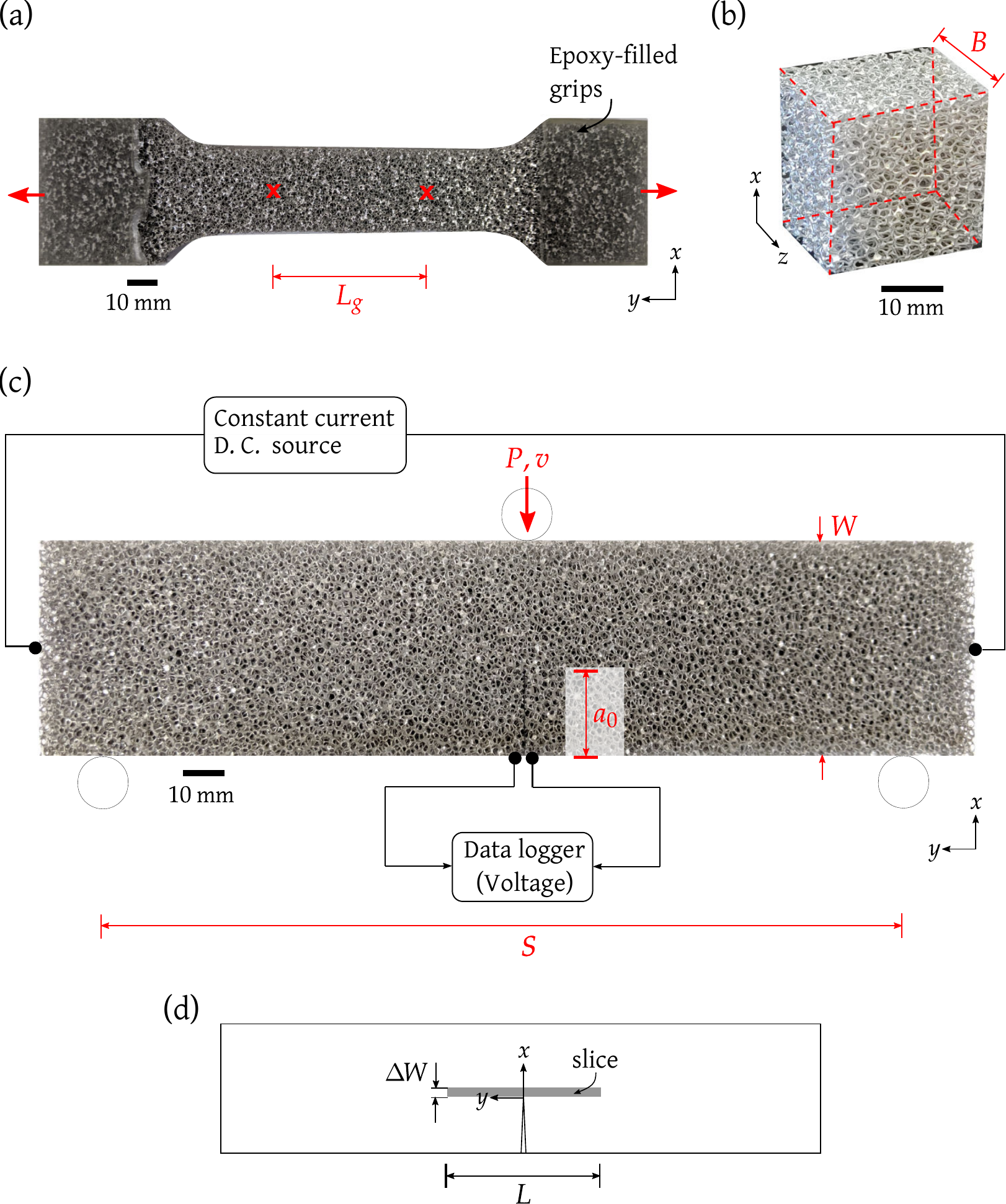}
\caption{Specimen geometries for \textbf{(a)} uniaxial tension test, \textbf{(b)} uniaxial compression test, and \textbf{(c)} fracture toughness test along with the apparatus. The out-of-plane thickness of all specimens is $B=\SI{26.4}{mm}$. \textbf{(d)} Definition of a slice used in the XCT analysis for damage visualisation.} 
\label{fig:specimens}
\end{figure}

%------------------------------------------------------------------------------------------------------
% MEASURED TENSION AND COMPRESSION STRESS STRAIN CURVES
%------------------------------------------------------------------------------------------------------
\newpage
\begin{figure}[H]
\includegraphics[scale=1]{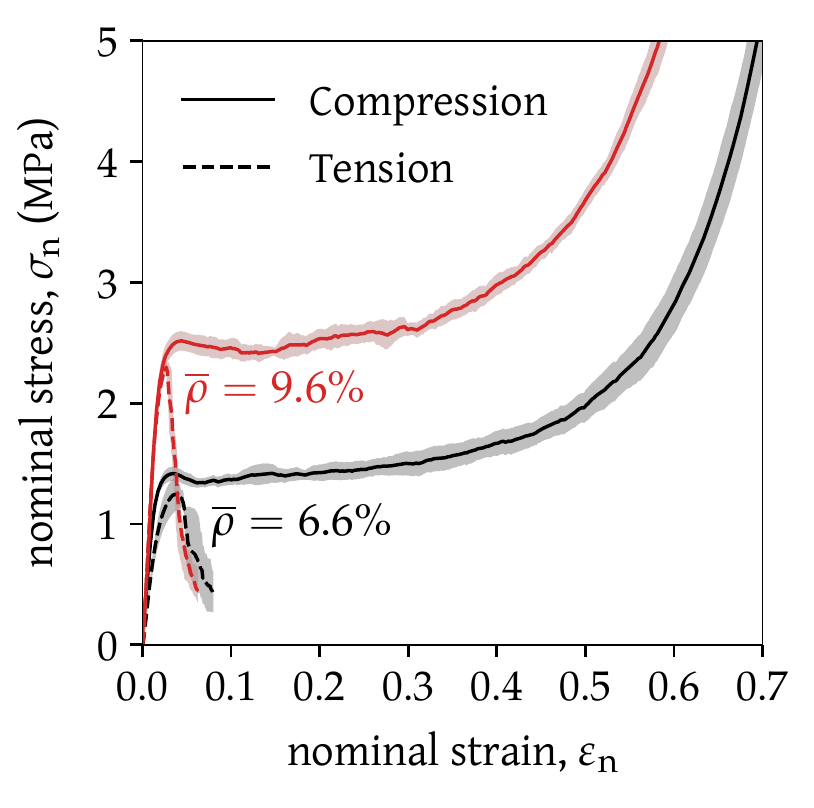}
\caption{Uniaxial tension, and in-plane compression, stress versus strain curves for the foams.}
\label{fig:exp_tension_compression}
\end{figure}

%------------------------------------------------------------------------------------------------------
% FRACTURE TEST - LOAD, CRACK EXTENSION, J-DELTA A CURVE, PLASTIC ZONE AT PEAK LOAD
%------------------------------------------------------------------------------------------------------
\newpage
\begin{figure}[H]
  \setlength{\labelsep}{-0.85cm}
  \sidesubfloat[]{\includegraphics[scale=1]{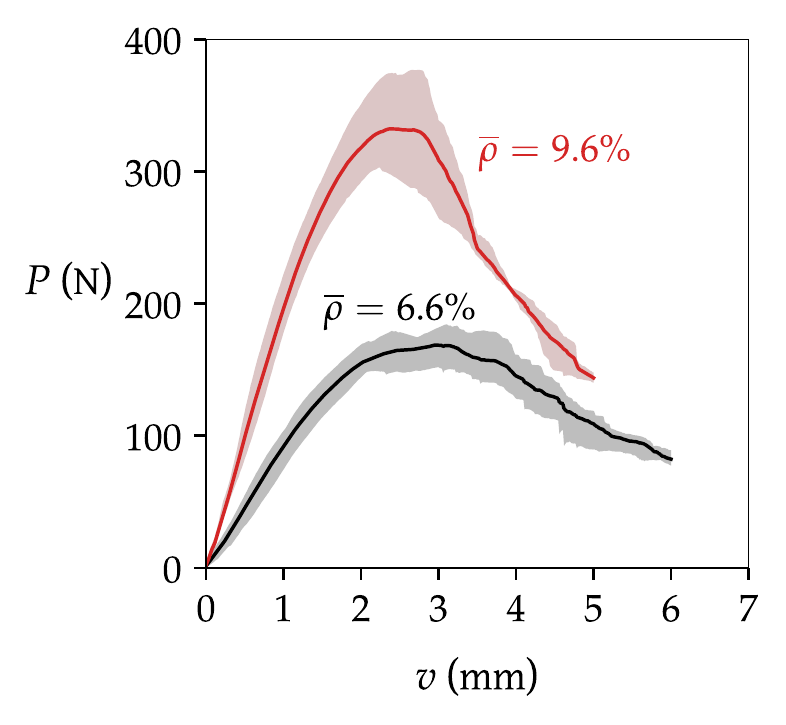} \label{fig:exp_frac_pv}}\hfill
  \setlength{\labelsep}{-0.85cm}
  \sidesubfloat[]{\includegraphics[scale=1]{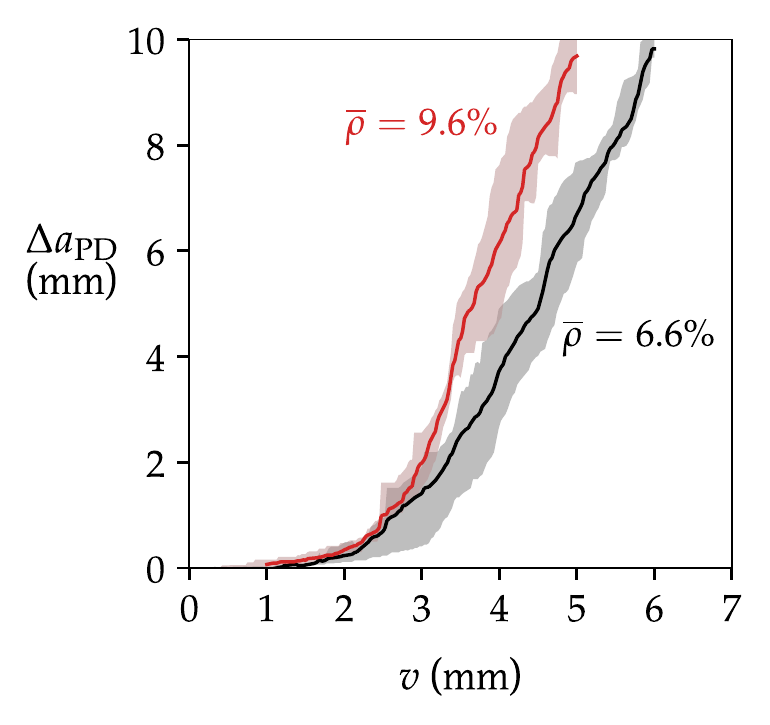}\label{fig:exp_frac_av}}\\[\baselineskip]
  \setlength{\labelsep}{-1.3cm}
  \sidesubfloat[]{\includegraphics[scale=1.0]{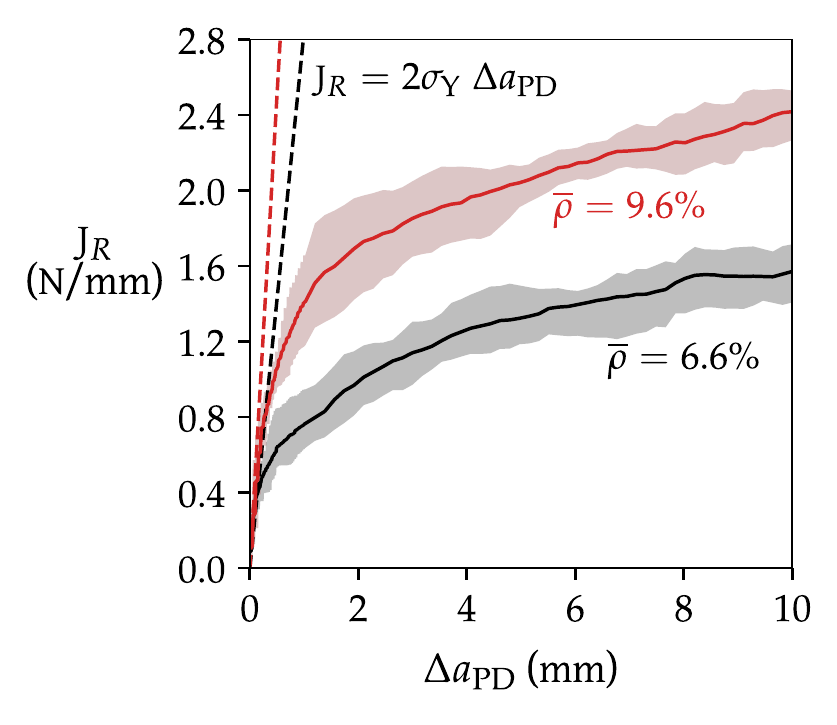}\label{fig:exp_frac_ja}}\hfill
  \setlength{\labelsep}{0.0cm}
  \sidesubfloat[]{\includegraphics[scale=1.0]{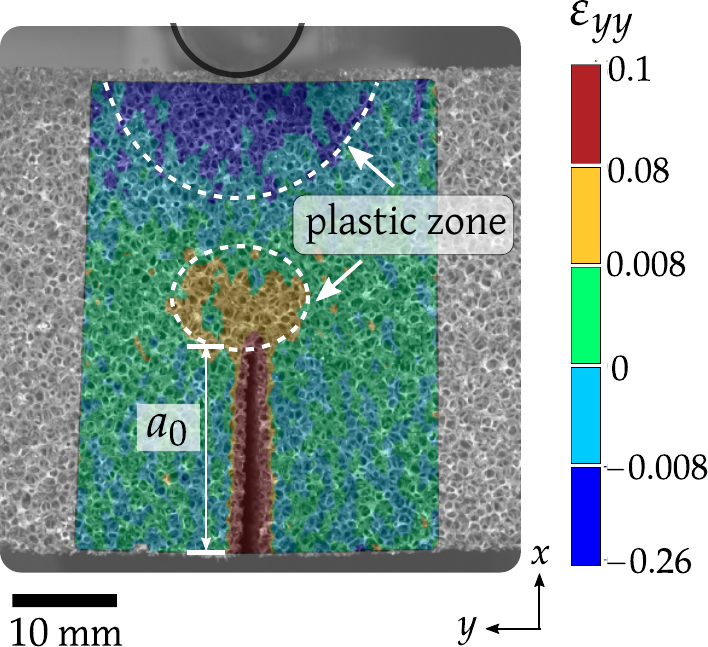}\label{fig:exp_frac_pzone}}
  \caption{Fracture response of deep-notched bend specimens: \textbf{(a)} Load $P$ versus cross-head displacement $v$ response, \textbf{(b)} crack extension $\Delta{a}_{\rm PD}$, as measured from the DCPD method, versus $v$,  \textbf{(c)} crack growth resistance curves for $\overline\rho=6.6\%$ and $9.6\%$ specimens, and \textbf{(d)} DIC contours of longitudinal strain $\varepsilon_{yy}$ in a specimen of $\overline\rho=6.6\%$ at peak load. The yield strain of this foam is $\varepsilon_{\rm Y}=\sigma_{\rm Y}/E = 0.008$.}
\label{fig:expt_senb}
\end{figure}

%------------------------------------------------------------------------------------------------------
% DAMAGE VISUALISATION USING CT - FRACTURE PATTERN AND DAMAGE VARIABLE F
%------------------------------------------------------------------------------------------------------
\newpage
\begin{figure}[H]
  \setlength{\labelsep}{-0.7cm}
  \sidesubfloat[]{\includegraphics[width=0.6\textwidth]{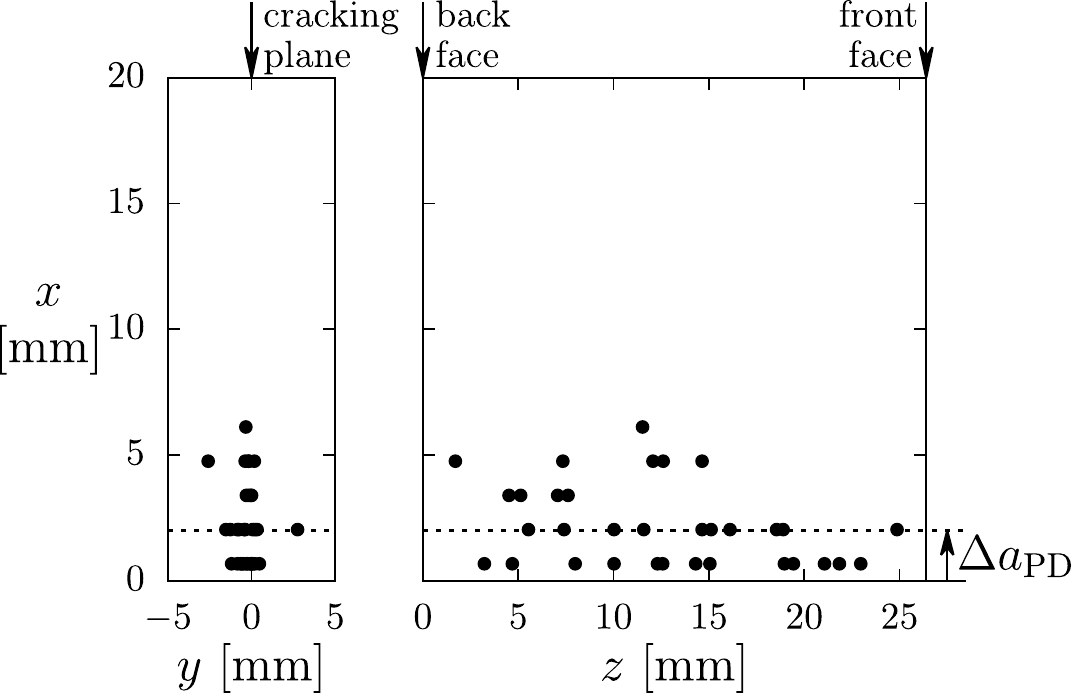}\label{fig:ERG_RelDen0066_Da2_failedStrutsCombinedZY_0066}}\\[\baselineskip]
  \setlength{\labelsep}{-0.7cm}
  \sidesubfloat[]{\includegraphics[width=0.515\textwidth]{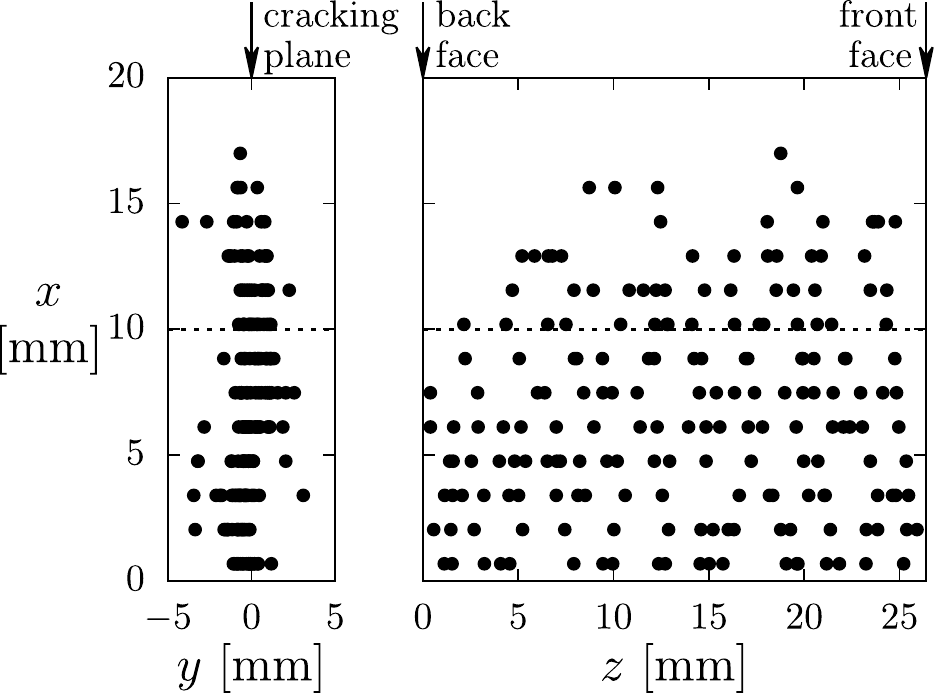}\label{fig:ERG_RelDen0066_Da10_failedStrutsCombinedZY_0066_1}}\hfill
  \setlength{\labelsep}{-0.7cm}
  \sidesubfloat[]{\includegraphics[width=0.475\textwidth]{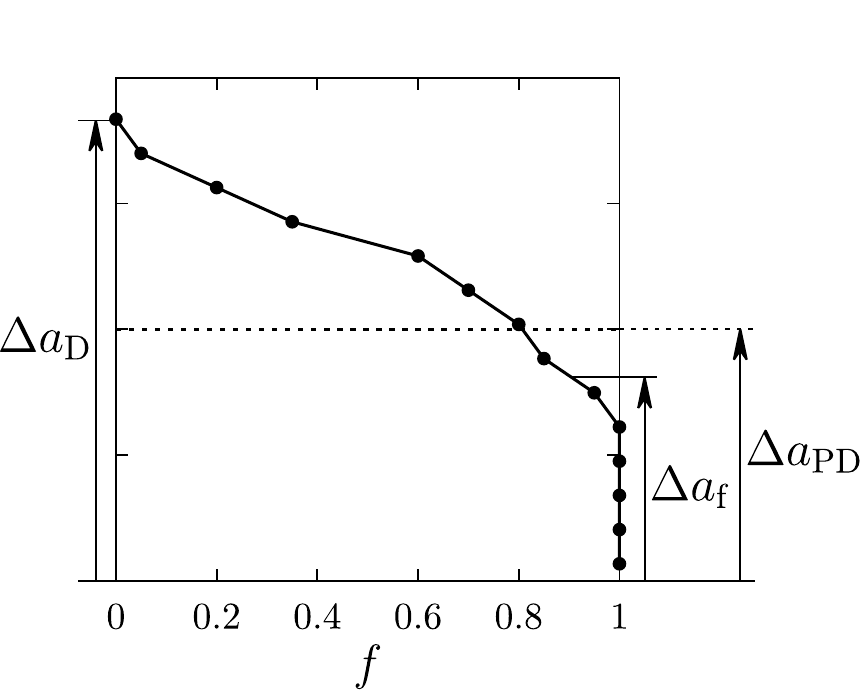}\label{fig:ERG_RelDen0066_Da10_failedStrutsCombinedZY_0066_2}}
  \caption{Distribution of failed struts as determined from XCT analysis in a representative specimen of $\overline\rho=6.6\%$: \subref{fig:ERG_RelDen0066_Da2_failedStrutsCombinedZY_0066} Location of failed struts for $\Delta a_\mathrm{PD}=2\,\mathrm{mm}$: the projected view in the $(x,y)$ plane shows failed struts over all $z$; likewise, the view in the $(x,z)$ plane shows failed struts over all $y-$values. 
    \subref{fig:ERG_RelDen0066_Da10_failedStrutsCombinedZY_0066_1}
    Location of failed struts for $\Delta a_\mathrm{PD}=10\,\mathrm{mm}$: again, the projected view in the $(x,y)$ is over all $z$; likewise, the view in the $(x,z)$ plane is over all $y$. 
    \subref{fig:ERG_RelDen0066_Da10_failedStrutsCombinedZY_0066_2} fraction of failed struts $f$ along the ligament ($x$-direction) for $\Delta a_\mathrm{PD}=10\,\mathrm{mm}$.
    }
  \label{fig:failedStrutsCombinedZY}
\end{figure}

%------------------------------------------------------------------------------------------------------
% FREE CRACK EXTENSION VERSUS PROCESS ZONE SIZE VERSUS PLASTIC ZONE SIZE - FROM CT
%------------------------------------------------------------------------------------------------------
\newpage
\begin{figure}[H]
  \setlength{\labelsep}{0.0cm}
  \sidesubfloat[]{\includegraphics[width=0.43\textwidth]{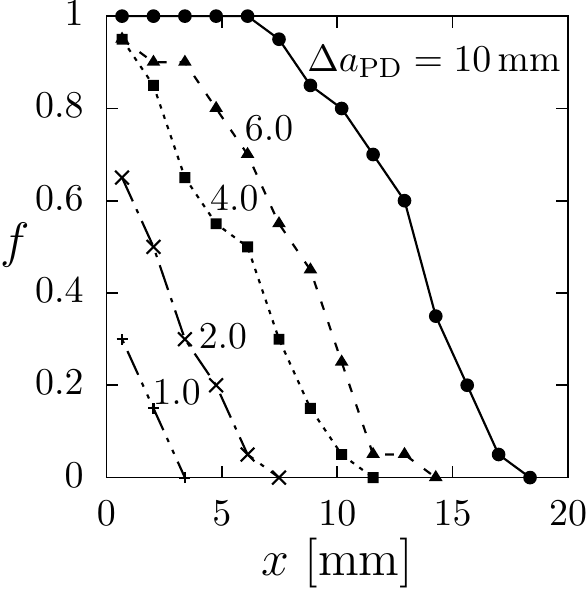}\label{fig:ERG_RelDen0066_2_failedStruts_stat}}\hfill
    \setlength{\labelsep}{-0.7cm}
  \sidesubfloat[]{\includegraphics[width=0.46\textwidth]{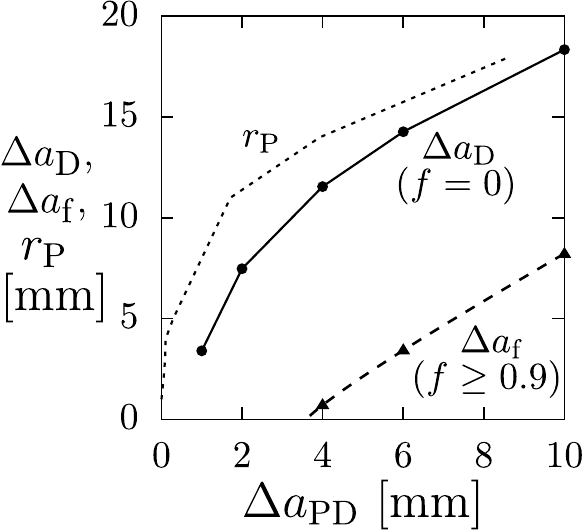}\label{fig:ERG_RelDen0066_FreeAndDamageCrackTip}}\\[\baselineskip]
  \setlength{\labelsep}{0.0cm}
  \sidesubfloat[]{\includegraphics[width=0.42\textwidth]{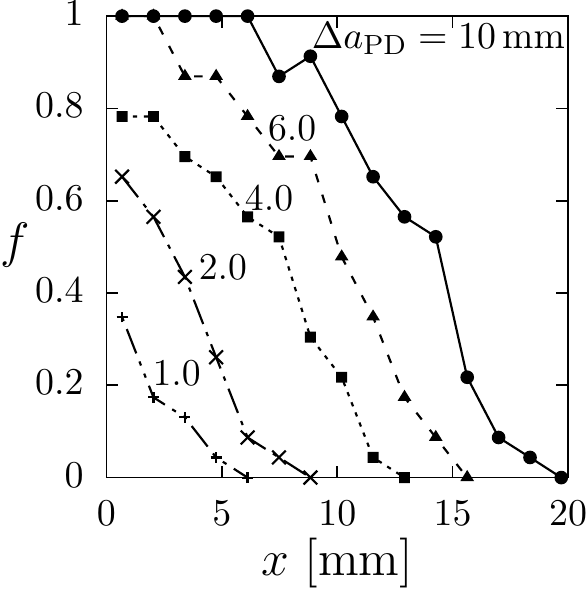}\label{fig:ERG_RelDen0096_2_failedStruts_stat}}\hfill
  \setlength{\labelsep}{-0.7cm}
  \sidesubfloat[]{\includegraphics[width=0.46\textwidth]{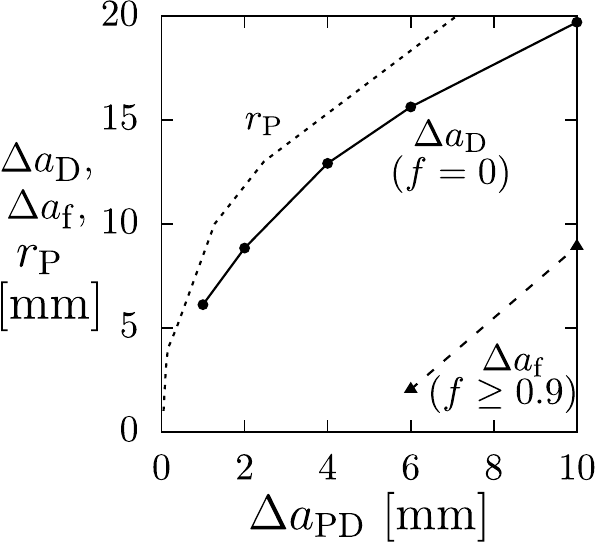}\label{fig:ERG_RelDen0096_FreeAndDamageCrackTip}}
  \caption{Extent of crack bridging for $\overline\rho=6.6\%$: \subref{fig:ERG_RelDen0066_2_failedStruts_stat} $f$ versus distance $x$ ahead of initial crack tip,  \subref{fig:ERG_RelDen0066_FreeAndDamageCrackTip}  traction-free crack extension $\Delta{a}_{\rm f}$, extent of damage zone $\Delta{a}_{\rm D}$, and extent of plastic zone size $r_{\rm P}$ versus inferred crack extension $\Delta{a}_{\rm PD}$. Extent of crack bridging for $\overline\rho=9.6\%$: \subref{fig:ERG_RelDen0096_2_failedStruts_stat}  $f$ versus distance $x$ ahead of initial crack tip,  \subref{fig:ERG_RelDen0096_FreeAndDamageCrackTip} $\Delta{a}_{\rm f}$, $\Delta{a}_{\rm D}$, and  $r_{\rm P}$ versus $\Delta{a}_{\rm PD}$. }
  \label{fig:damagePlasticZone}
\end{figure}

%------------------------------------------------------------------------------------------------------
% FE MODEL OF SENB TEST
%------------------------------------------------------------------------------------------------------
\newpage
\begin{figure}[H]
{\includegraphics[scale=1.0]{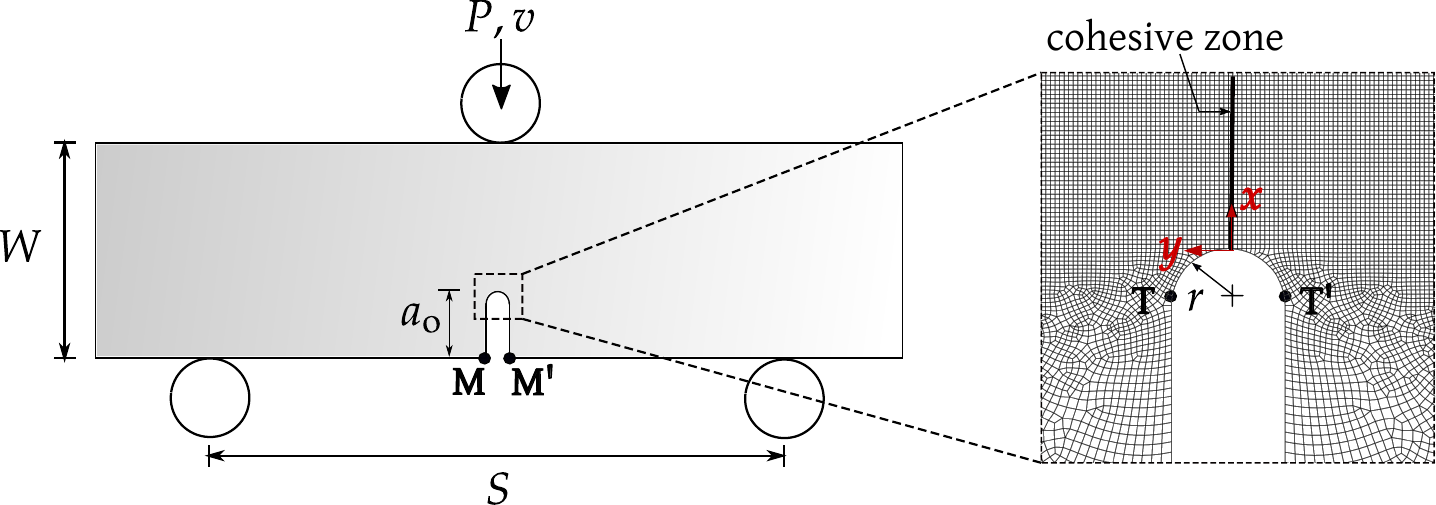}}
\caption{Geometry of the SENB specimen used in FE simulations and details of the notch tip mesh.}
\label{fig:fe_senb_geometry}
\end{figure}
\vspace{10mm}

%------------------------------------------------------------------------------------------------------
% ASSUMED MATERIAL RESPONSE ** VALIDATION  WITH NO-NOTCH CASE
%------------------------------------------------------------------------------------------------------
\begin{figure}[H]
  \setlength{\labelsep}{-0.3cm}
  \sidesubfloat[]{\includegraphics[scale=1]{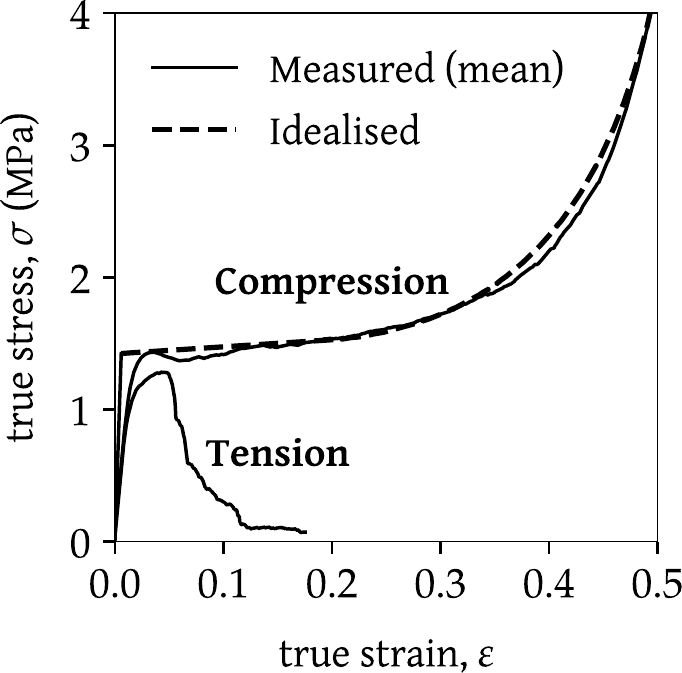} \label{fig:fe_material}}\hfill
  \setlength{\labelsep}{-1.1cm}
  \sidesubfloat[]{\includegraphics[scale=1]{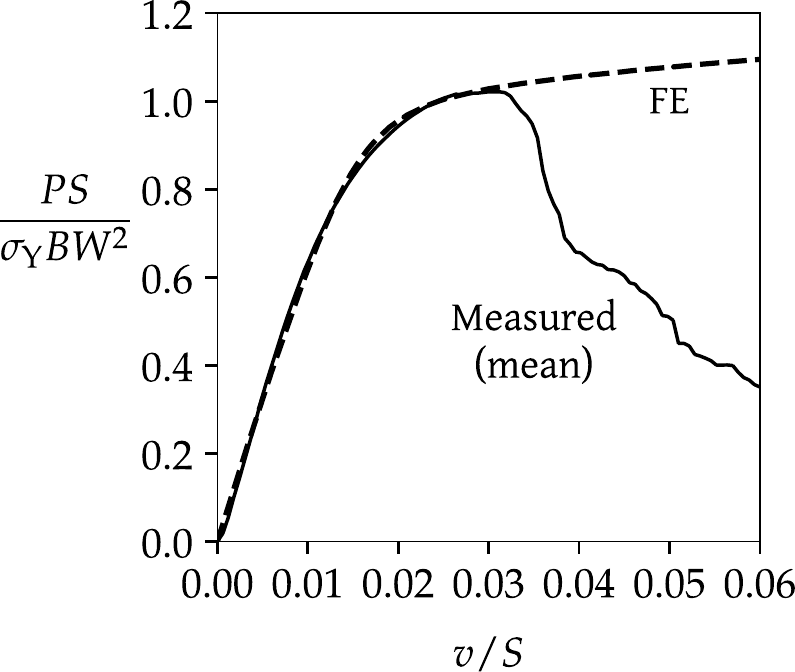}\label{fig:fe_bend}}
\caption{\textbf{(a)} Idealised true stress versus true strain response for the foam of $\overline\rho=6.6\%$. \textbf{(b)} Comparison of the predicted and measured load versus displacement response of the bend specimen (without a pre-notch).}
\label{fig:fe_foamproperties}
\end{figure}

%------------------------------------------------------------------------------------------------------
% ASSUMED SHAPE OF COHESIVE LAW ** BEND RESPONSE FOR LIMITING CASES OF COHESIVE PARAMETERS
%------------------------------------------------------------------------------------------------------
\newpage
\begin{figure}[H]
  \setlength{\labelsep}{-0.4cm}
  \sidesubfloat[]{\includegraphics[scale=0.98]{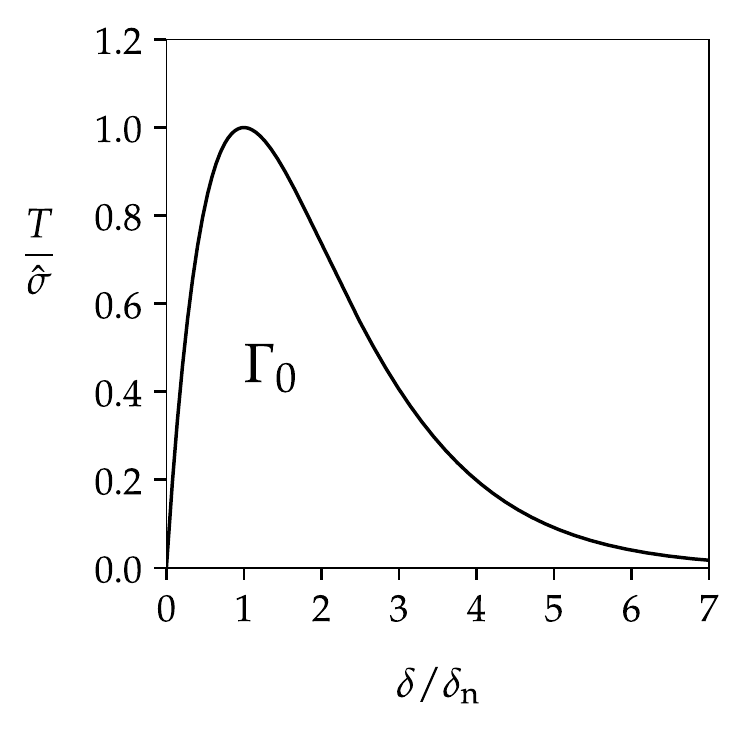}\label{fig:coh-law}}\hfill
  \setlength{\labelsep}{-1.1cm}
  \sidesubfloat[]{\includegraphics[scale=0.98]{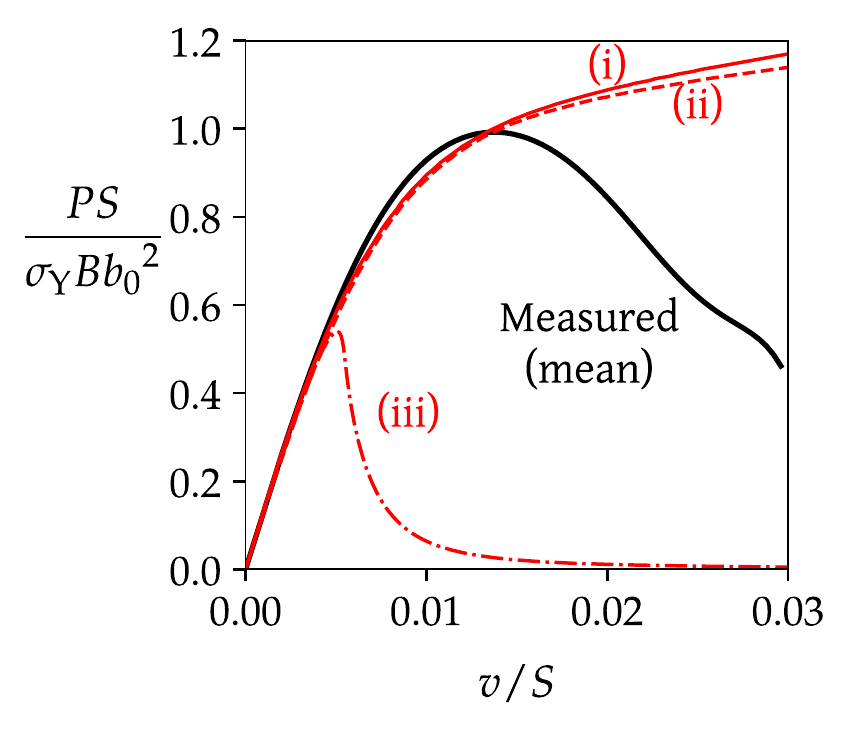}\label{fig:limitingcases}}
  \caption{ \textbf{(a)} Assumed traction $T$ versus separation $\delta$ response for the cohesive zone.  \textbf{(b)} Predictions of the load versus displacement response of a deeply notched bend specimen with the 3 limiting cases of cohesive properties: (i) $\hat\sigma=\Gamma_0=\infty$, (ii) rigid, perfectly plastic cohesive zone $\left( \hat\sigma=\sigma_{\rm Y}, \Gamma_0=\infty \right)$, and (iii) $\hat\sigma=\sigma_{\rm Y}, \Gamma_0 \rightarrow 0$. The measured mean response for $\overline\rho=6.6\%$ is included.}
  % \label{fig:fe_senb}
\end{figure}
\vspace{10mm}
  
%------------------------------------------------------------------------------------------------------
% GOODNESS-OF-FIT CONTOURS ** COHESIVE LAWS FOR 3 BEST-FIT CASES
%------------------------------------------------------------------------------------------------------
\begin{figure}[H]
  \setlength{\labelsep}{-0.5cm}
  \sidesubfloat[]{\includegraphics[scale=1]{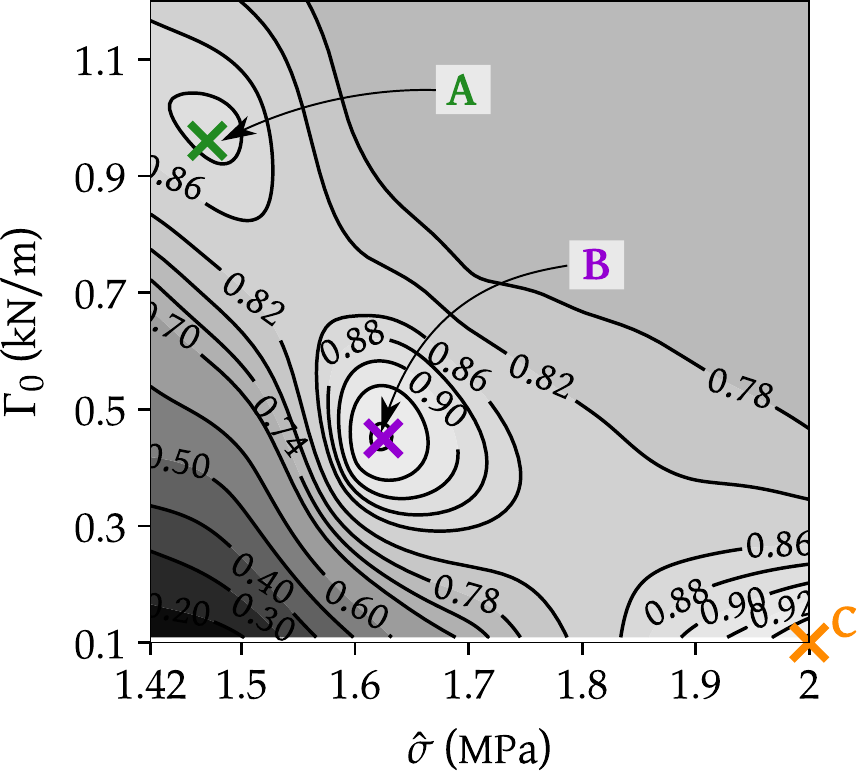}\label{fig:contours_chi}}
  \sidesubfloat[]{\includegraphics[scale=1]{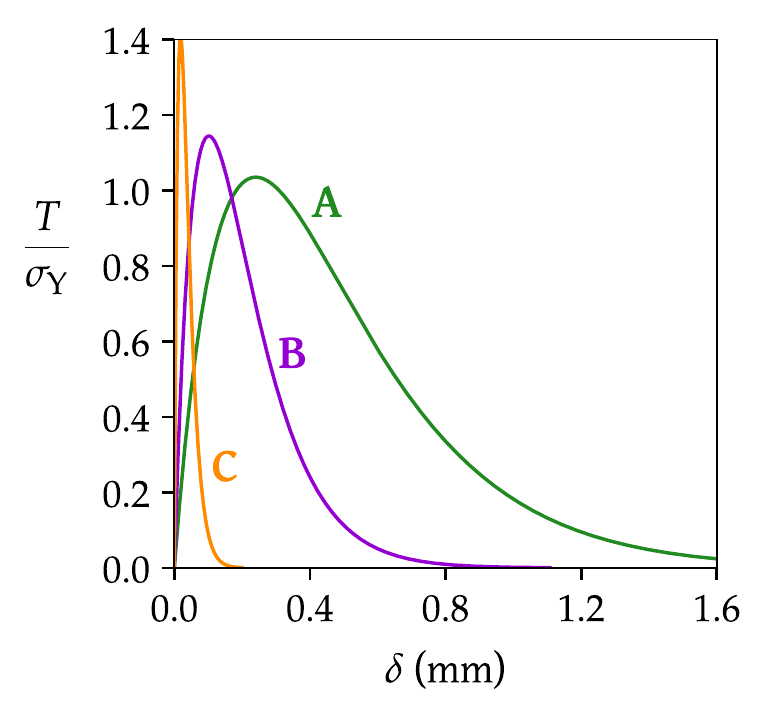}\label{fig:bestfit_cohlaw}}
  \caption{\textbf{(a)}  Contours of goodness-of-fit $\chi$, and \textbf{(b)} cohesive laws for the 3 best fitting cases.}
  \label{fig:contours_limitingcases}
\end{figure}

%------------------------------------------------------------------------------------------------------
% FE BEST-FIT CASES VERSUS EXPERIMENT
%------------------------------------------------------------------------------------------------------
\newpage
\begin{figure}[H]
\setlength{\labelsep}{-1.5cm}
\sidesubfloat[]{\includegraphics[scale=1]{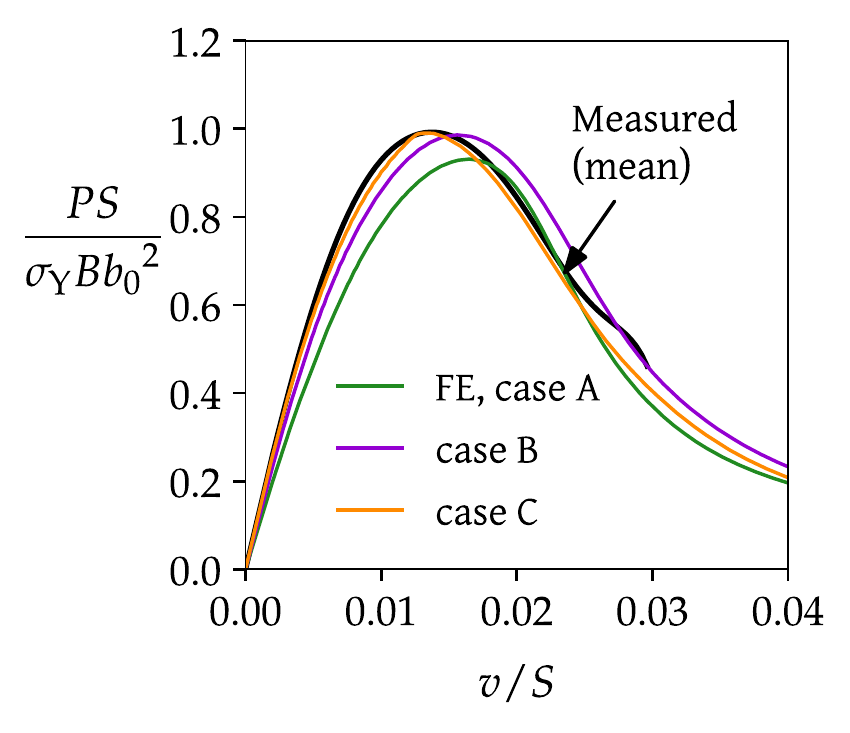}\label{fig:bestfit_load}}
\setlength{\labelsep}{-0.75cm}
\sidesubfloat[]{\includegraphics[scale=1]{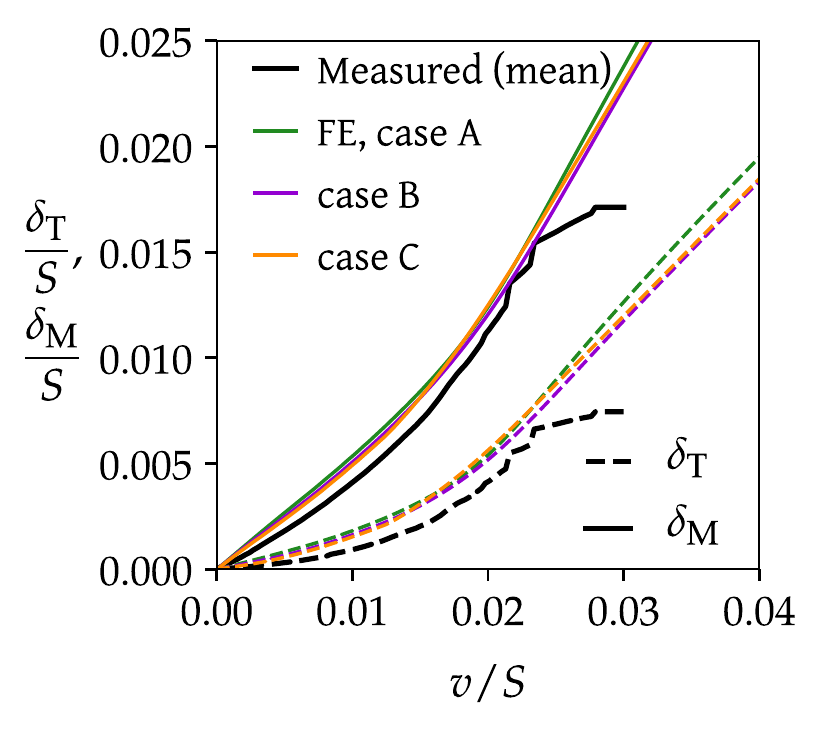}\label{fig:bestfit_cmod}}
\vspace{3mm}
\setlength{\labelsep}{-0.85cm}
\sidesubfloat[]{\includegraphics[scale=1]{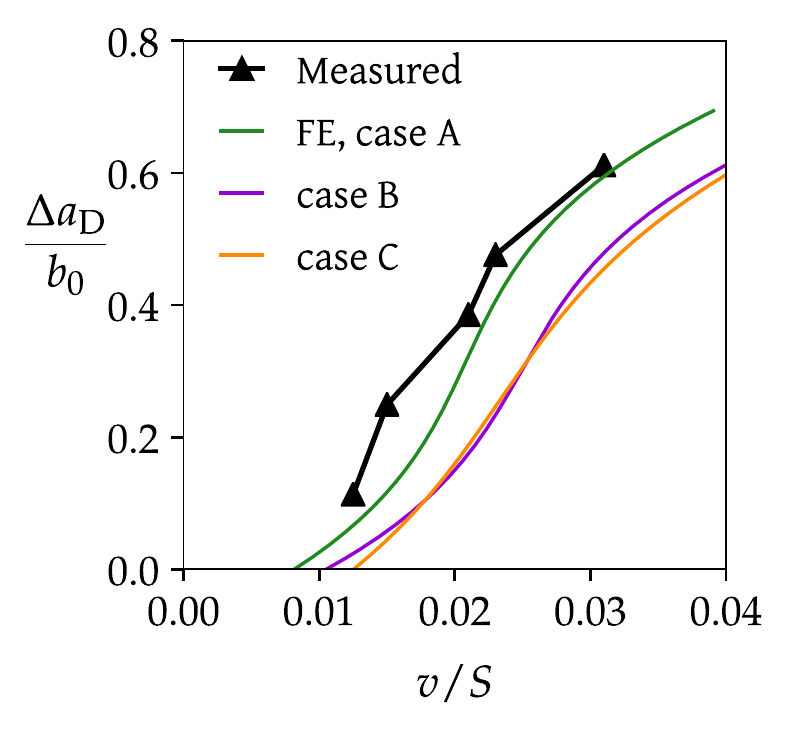}\label{fig:bestfit_dadamage}}\hfill
\sidesubfloat[]{\includegraphics[scale=1]{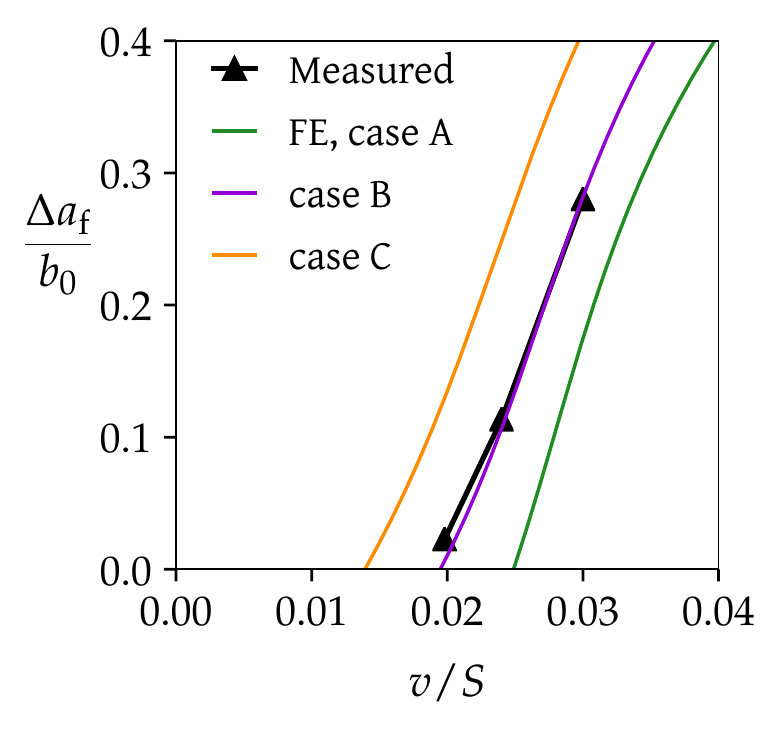}\label{fig:bestfit_da}}
\vspace{3mm}
\sidesubfloat[]{\includegraphics[scale=1]{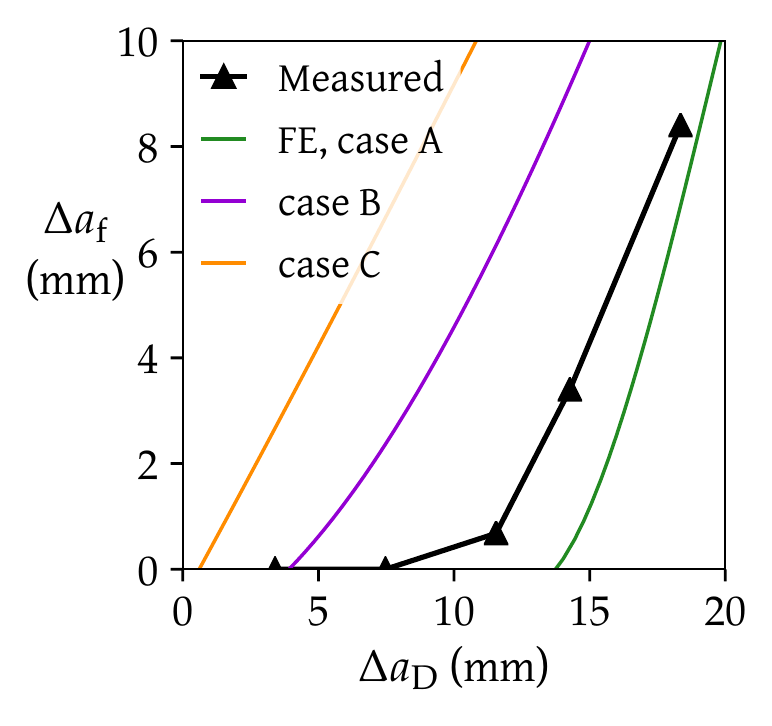}\label{fig:bestfit_dafreedamage}}\hfill
\setlength{\labelsep}{-0.5cm}
\sidesubfloat[]{\includegraphics[scale=1]{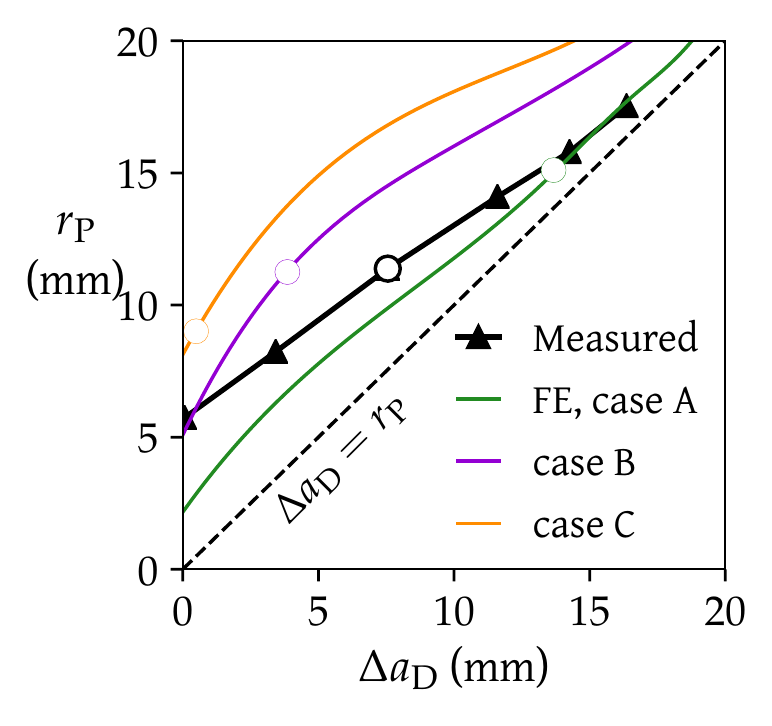}\label{fig:bestfit_pzdadamage}}
\vspace{2mm}
\caption{Comparison of the FE predictions for cases A, B, and C labelled in \cref{fig:contours_chi} with experimental observations for $\overline\rho=6.6\%$: \textbf{(a)} load $P$ versus displacement $v$ response, \textbf{(b)} evolution of crack tip opening displacement $(\delta_{\rm T})$ and crack mouth opening displacement $(\delta_{\rm M})$ with $v$, \textbf{(c)} extent of the damage/bridging zone $\Delta{a}_{\rm D}$ versus displacement $v$, \textbf{(d)}  traction-free crack extension $\Delta{a}_{\rm f}$ versus $v$, \textbf{(e)} $\Delta{a}_{\rm D}$ versus  $\Delta{a}_{\rm f}$, and \textbf{(f)}  relative extent of the plastic zone $r_{\rm P}$ to the damage zone (with $\textsf{O}$ denoting the point of initiation of a traction-free crack,  $\Delta{a}_{\rm f}=0^+$).}
\label{fig:fe_bestfit}
\end{figure}

%------------------------------------------------------------------------------------------------------
% DENT TESTS: EXPERIMENT VERSUS PREDICTION, LOAD
%------------------------------------------------------------------------------------------------------
\newpage
\begin{figure}[H]
\sidesubfloat[]{\includegraphics[scale=1]{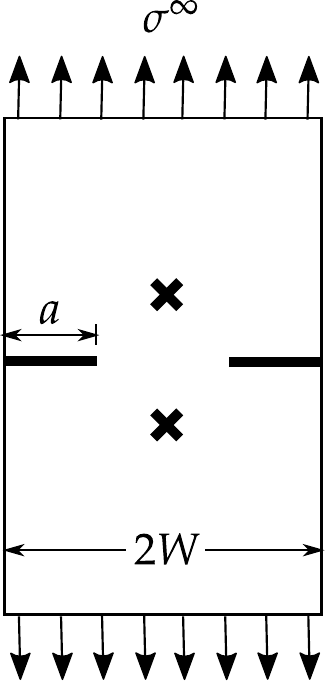}\label{fig:dent_geometry}}
\hspace{10mm}
  \setlength{\labelsep}{-0.75cm}
  \sidesubfloat[]{\includegraphics[scale=1]{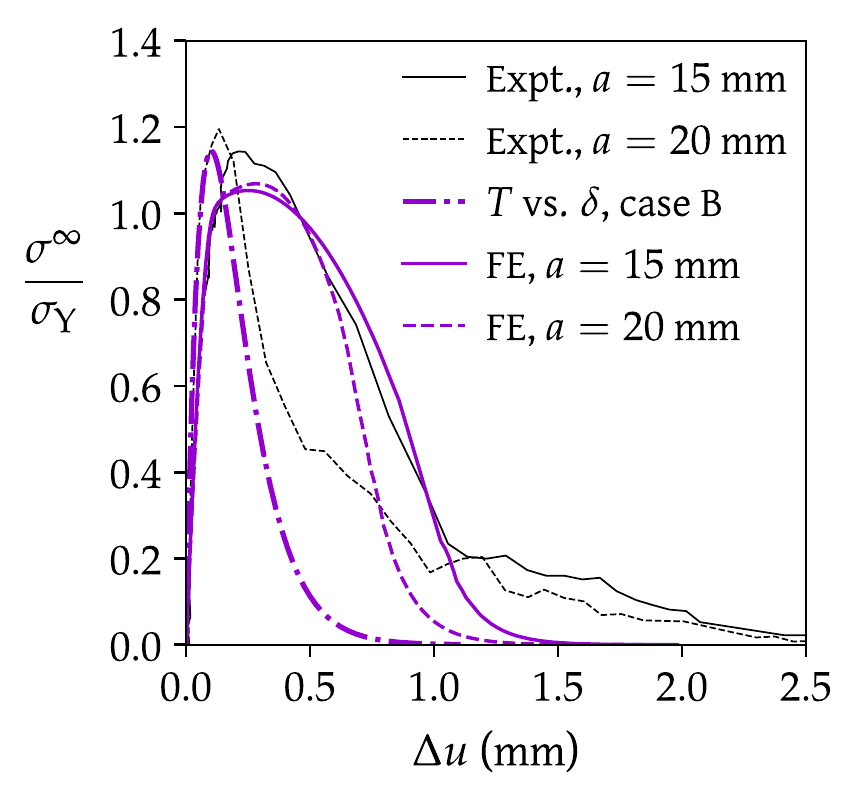}\label{fig:dent_load}}
\vspace{5mm}
\caption{Response of a double edge notch specimen under tension: \textbf{(a)} geometry and loading, \textbf{(b)} measured and predicted load versus displacement response. The cohesive traction $T$ versus opening $\delta$ response for the best fitting case, case B, of the bend toughness test is included.}
\label{fig:dent_exp_fe_load}
\end{figure}

%------------------------------------------------------------------------------------------------------
% DENT TESTS: EXPERIMENT VERSUS PREDICTION, DISPLACEMENT AND STRAIN CONTOURS
%------------------------------------------------------------------------------------------------------
\newpage
\begin{figure}[H]
\includegraphics[scale=1]{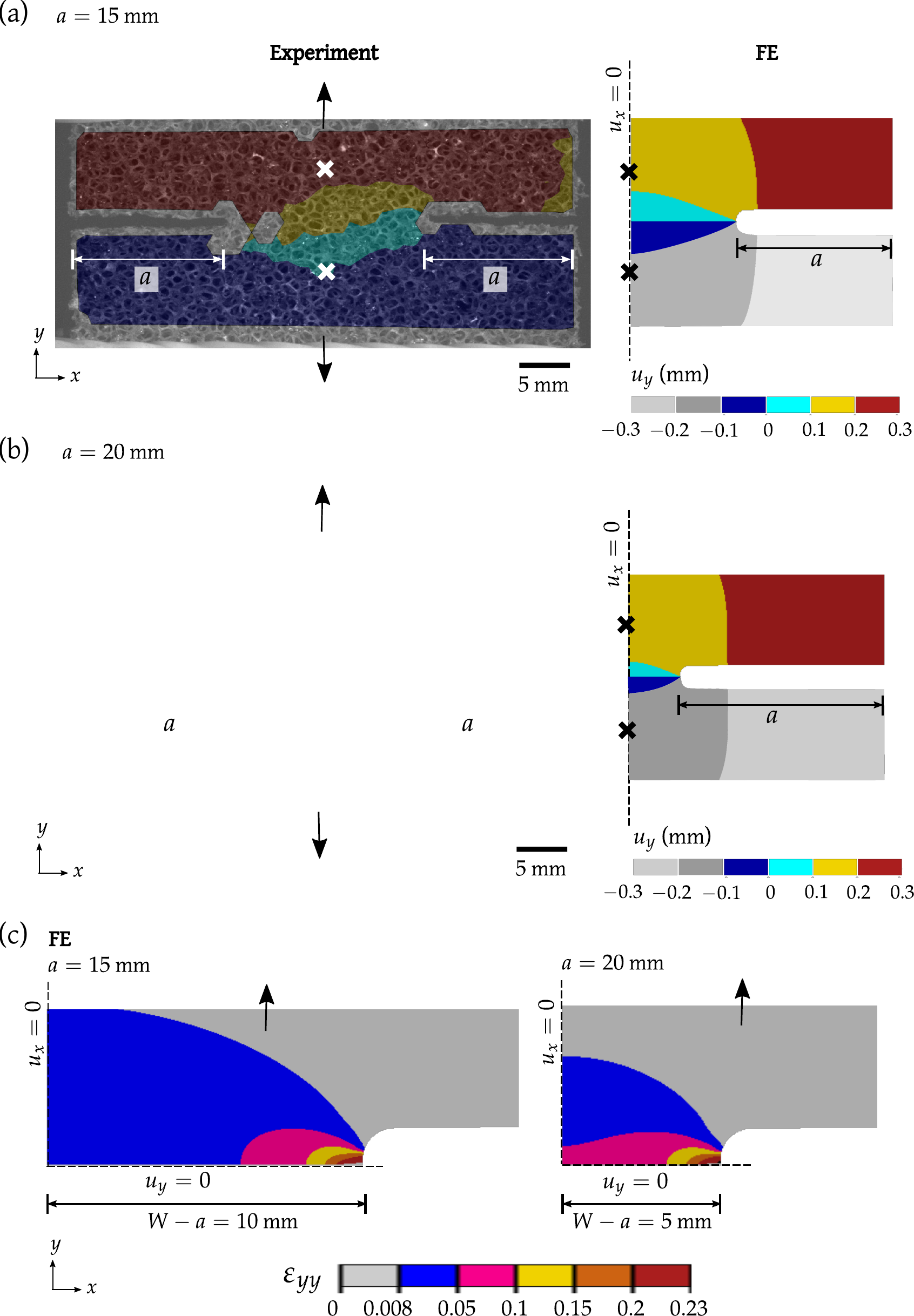}
\caption{Comparison of the displacement contours from DIC and FE simulations for \textbf{(a)} $a=15$ mm and \textbf{(b)} $a=20$ mm at peak load. \textbf{(c)} FE contours of strain in the loading direction for the two notch geometries, also at peak load.}
\label{fig:dent_u1}
\end{figure}

%% file: Main.bbl
\begin{thebibliography}{33}
\providecommand{\natexlab}[1]{#1}
\providecommand{\url}[1]{\texttt{#1}}
\providecommand{\urlprefix}{URL }
\expandafter\ifx\csname urlstyle\endcsname\relax
  \providecommand{\doi}[1]{doi:\discretionary{}{}{}#1}\else
  \providecommand{\doi}[1]{doi:\discretionary{}{}{}\begingroup
  \urlstyle{rm}\url{#1}\endgroup}\fi
\providecommand{\bibinfo}[2]{#2}

\bibitem[{{ASTM E1820}(2013)}]{e1820}
\bibinfo{author}{{ASTM E1820}}, \bibinfo{title}{Standard test method for
  measurement of fracture toughness}, \bibinfo{journal}{American Society for
  Testing and Materials West Conshohocken, PA} .

\bibitem[{Jelitto and Schneider(2019)}]{Jelitto2019}
\bibinfo{author}{H.~Jelitto}, \bibinfo{author}{G.~A. Schneider},
  \bibinfo{title}{Fracture toughness of porous materials – Experimental
  methods and data}, \bibinfo{journal}{Data in Brief} \bibinfo{volume}{23},
  \doi{\bibinfo{doi}{10.1016/j.dib.2019.103709}}.

\bibitem[{McCullough et~al.(1999)McCullough, Fleck, and Ashby}]{McCullough1999}
\bibinfo{author}{K.~Y.~G. McCullough}, \bibinfo{author}{N.~A. Fleck},
  \bibinfo{author}{M.~F. Ashby}, \bibinfo{title}{Toughness of aluminum alloy
  foams}, \bibinfo{journal}{Acta Materialia}
  \bibinfo{volume}{47}~(\bibinfo{number}{8}) (\bibinfo{year}{1999})
  \bibinfo{pages}{2331--2343},
  \doi{\bibinfo{doi}{10.1016/S1359-6454(99)00125-1}}.

\bibitem[{Kashef et~al.(2010)Kashef, Asgari, Hilditch, Yan, Goel, and
  Hodgson}]{Kashef2010}
\bibinfo{author}{S.~Kashef}, \bibinfo{author}{A.~Asgari},
  \bibinfo{author}{T.~B. Hilditch}, \bibinfo{author}{W.~Yan},
  \bibinfo{author}{V.~K. Goel}, \bibinfo{author}{P.~D. Hodgson},
  \bibinfo{title}{Fracture toughness of titanium foams for medical
  applications}, \bibinfo{journal}{Materials Science and Engineering: A}
  \bibinfo{volume}{527}~(\bibinfo{number}{29}) (\bibinfo{year}{2010})
  \bibinfo{pages}{7689 -- 7693},
  \doi{\bibinfo{doi}{10.1016/j.msea.2010.08.044}}.

\bibitem[{Combaz and Mortensen(2010)}]{Combaz2010}
\bibinfo{author}{E.~Combaz}, \bibinfo{author}{A.~Mortensen},
  \bibinfo{title}{Fracture toughness of Al replicated foam},
  \bibinfo{journal}{Acta Materialia}
  \bibinfo{volume}{58}~(\bibinfo{number}{14}) (\bibinfo{year}{2010})
  \bibinfo{pages}{4590--4603},
  \doi{\bibinfo{doi}{10.1016/j.actamat.2010.04.025}}.

\bibitem[{Olurin et~al.(2000)Olurin, Fleck, and Ashby}]{Olurin2000}
\bibinfo{author}{O.~B. Olurin}, \bibinfo{author}{N.~A. Fleck},
  \bibinfo{author}{M.~F. Ashby}, \bibinfo{title}{Deformation and fracture of
  aluminium foams}, \bibinfo{journal}{Materials Science and Engineering A}
  \bibinfo{volume}{291}~(\bibinfo{number}{1-2}) (\bibinfo{year}{2000})
  \bibinfo{pages}{136--146},
  \doi{\bibinfo{doi}{10.1016/S0921-5093(00)00954-0}}.

\bibitem[{Motz and Pippan(2001)}]{Motz2001}
\bibinfo{author}{C.~Motz}, \bibinfo{author}{R.~Pippan},
  \bibinfo{title}{Deformation behaviour of closed-cell aluminium foams in
  tension}, \bibinfo{journal}{Acta Materialia}
  \bibinfo{volume}{49}~(\bibinfo{number}{13}) (\bibinfo{year}{2001})
  \bibinfo{pages}{2463--2470},
  \doi{\bibinfo{doi}{10.1016/S1359-6454(01)00152-5}}.

\bibitem[{O'Masta et~al.(2017)O'Masta, Dong, St-Pierre, Wadley, and
  Deshpande}]{Masta2017}
\bibinfo{author}{M.~R. O'Masta}, \bibinfo{author}{L.~Dong},
  \bibinfo{author}{L.~St-Pierre}, \bibinfo{author}{H.~N.~G. Wadley},
  \bibinfo{author}{V.~S. Deshpande}, \bibinfo{title}{The fracture toughness of
  octet-truss lattices}, \bibinfo{journal}{Journal of the Mechanics and Physics
  of Solids} \bibinfo{volume}{98} (\bibinfo{year}{2017})
  \bibinfo{pages}{271--289}, \doi{\bibinfo{doi}{10.1016/j.jmps.2016.09.009}}.

\bibitem[{S{\o}rensen and Jacobsen(1998)}]{Sorensen1998}
\bibinfo{author}{B.~F. S{\o}rensen}, \bibinfo{author}{T.~K. Jacobsen},
  \bibinfo{title}{Large-scale bridging in composites: R-curves and bridging
  laws}, \bibinfo{journal}{Composites Part A: Applied Science and
  Manufacturing} \bibinfo{volume}{29}~(\bibinfo{number}{11})
  (\bibinfo{year}{1998}) \bibinfo{pages}{1443--1451},
  \doi{\bibinfo{doi}{10.1016/S1359-835X(98)00025-6}}.

\bibitem[{Zok and Hom(1990)}]{Zok1990}
\bibinfo{author}{F.~W. Zok}, \bibinfo{author}{C.~L. Hom}, \bibinfo{title}{Large
  scale bridging in brittle matrix composites}, \bibinfo{journal}{Acta
  Metallurgica et Materialia} \bibinfo{volume}{38}~(\bibinfo{number}{10})
  (\bibinfo{year}{1990}) \bibinfo{pages}{1895--1904},
  \doi{\bibinfo{doi}{10.1016/0956-7151(90)90301-V}}.

\bibitem[{Deshpande and Fleck(2000)}]{Deshpande2000}
\bibinfo{author}{V.~S. Deshpande}, \bibinfo{author}{N.~A. Fleck},
  \bibinfo{title}{Isotropic constitutive models for metallic foams},
  \bibinfo{journal}{Journal of the Mechanics and Physics of Solids}
  \bibinfo{volume}{48}~(\bibinfo{number}{6}) (\bibinfo{year}{2000})
  \bibinfo{pages}{1253--1283},
  \doi{\bibinfo{doi}{10.1016/S0022-5096(99)00082-4}}.

\bibitem[{Teko\u{g}lu and Onck(2008)}]{Tekoglu2008}
\bibinfo{author}{C.~Teko\u{g}lu}, \bibinfo{author}{P.~R. Onck},
  \bibinfo{title}{Size effects in two-dimensional Voronoi foams: A comparison
  between generalized continua and discrete models}, \bibinfo{journal}{Journal
  of the Mechanics and Physics of Solids}
  \bibinfo{volume}{56}~(\bibinfo{number}{12}) (\bibinfo{year}{2008})
  \bibinfo{pages}{3541--3564}, \doi{\bibinfo{doi}{10.1016/j.jmps.2008.06.007}}.

\bibitem[{Dillard et~al.(2006)Dillard, Forest, and Ienny}]{Dillard2006}
\bibinfo{author}{T.~Dillard}, \bibinfo{author}{S.~Forest},
  \bibinfo{author}{P.~Ienny}, \bibinfo{title}{Micromorphic continuum modelling
  of the deformation and fracture behaviour of nickel foams},
  \bibinfo{journal}{European Journal of Mechanics, A/Solids}
  \bibinfo{volume}{25}~(\bibinfo{number}{3}) (\bibinfo{year}{2006})
  \bibinfo{pages}{526--549},
  \doi{\bibinfo{doi}{10.1016/j.euromechsol.2005.11.006}}.

\bibitem[{Teko\u{g}lu et~al.(2011)Teko\u{g}lu, Gibson, Pardoen, and
  Onck}]{Tekoglu2011}
\bibinfo{author}{C.~Teko\u{g}lu}, \bibinfo{author}{L.~J. Gibson},
  \bibinfo{author}{T.~Pardoen}, \bibinfo{author}{P.~R. Onck},
  \bibinfo{title}{Size effects in foams: Experiments and modeling},
  \bibinfo{journal}{Progress in Materials Science}
  \bibinfo{volume}{56}~(\bibinfo{number}{2}) (\bibinfo{year}{2011})
  \bibinfo{pages}{109--138},
  \doi{\bibinfo{doi}{10.1016/j.pmatsci.2010.06.001}}.

\bibitem[{Andrews and Gibson(2001)}]{Andrews2001}
\bibinfo{author}{E.~W. Andrews}, \bibinfo{author}{L.~J. Gibson},
  \bibinfo{title}{The influence of crack-like defects on the tensile strength
  of an open-cell aluminum foam}, \bibinfo{journal}{Scripta Materialia}
  \bibinfo{volume}{44}~(\bibinfo{number}{7}) (\bibinfo{year}{2001})
  \bibinfo{pages}{1005--1010},
  \doi{\bibinfo{doi}{10.1016/S1359-6462(01)00673-X}}.

\bibitem[{Onck(2001)}]{Onck2001}
\bibinfo{author}{P.~R. Onck}, \bibinfo{title}{Notch-strengthening in
  two-dimensional foams}, \bibinfo{journal}{Le Journal de Physique IV}
  \bibinfo{volume}{11}, \doi{\bibinfo{doi}{10.1051/jp4:2001526}}.

\bibitem[{Mangipudi and Onck(2011{\natexlab{a}})}]{Mangipudi2011a}
\bibinfo{author}{K.~R. Mangipudi}, \bibinfo{author}{P.~R. Onck},
  \bibinfo{title}{Notch sensitivity of ductile metallic foams: A computational
  study}, \bibinfo{journal}{Acta Materialia}
  \bibinfo{volume}{59}~(\bibinfo{number}{19})
  (\bibinfo{year}{2011}{\natexlab{a}}) \bibinfo{pages}{7356--7367},
  \doi{\bibinfo{doi}{10.1016/j.actamat.2011.07.071}}.

\bibitem[{Mangipudi and Onck(2011{\natexlab{b}})}]{Mangipudi2011b}
\bibinfo{author}{K.~R. Mangipudi}, \bibinfo{author}{P.~R. Onck},
  \bibinfo{title}{Multiscale modelling of damage and failure in two-dimensional
  metallic foams}, \bibinfo{journal}{Journal of the Mechanics and Physics of
  Solids} \bibinfo{volume}{59}~(\bibinfo{number}{7})
  (\bibinfo{year}{2011}{\natexlab{b}}) \bibinfo{pages}{1437--1461},
  \doi{\bibinfo{doi}{10.1016/j.jmps.2011.02.008}}.

\bibitem[{Seiler et~al.(2019{\natexlab{a}})Seiler, Tankasala, and
  Fleck}]{Seiler2019a}
\bibinfo{author}{P.~E. Seiler}, \bibinfo{author}{H.~C. Tankasala},
  \bibinfo{author}{N.~A. Fleck}, \bibinfo{title}{The role of defects in
  dictating the strength of brittle honeycombs made by rapid prototyping},
  \bibinfo{journal}{Acta Materialia} \bibinfo{volume}{171}
  (\bibinfo{year}{2019}{\natexlab{a}}) \bibinfo{pages}{190--200},
  \doi{\bibinfo{doi}{10.1016/j.actamat.2019.03.036}}.

\bibitem[{Seiler et~al.(2019{\natexlab{b}})Seiler, Tankasala, and
  Fleck}]{Seiler2019b}
\bibinfo{author}{P.~E. Seiler}, \bibinfo{author}{H.~C. Tankasala},
  \bibinfo{author}{N.~A. Fleck}, \bibinfo{title}{Creep failure of honeycombs
  made by rapid prototyping}, \bibinfo{journal}{Acta Materialia}
  \bibinfo{volume}{178} (\bibinfo{year}{2019}{\natexlab{b}})
  \bibinfo{pages}{122--134},
  \doi{\bibinfo{doi}{10.1016/j.actamat.2019.07.054}}.

\bibitem[{Tvergaard and Hutchinson(1992)}]{Tvergaard1992}
\bibinfo{author}{V.~Tvergaard}, \bibinfo{author}{J.~W. Hutchinson},
  \bibinfo{title}{The relation between crack growth resistance and fracture
  process parameters in elastic-plastic solids}, \bibinfo{journal}{Journal of
  the Mechanics and Physics of Solids}
  \bibinfo{volume}{40}~(\bibinfo{number}{6}) (\bibinfo{year}{1992})
  \bibinfo{pages}{1377--1397},
  \doi{\bibinfo{doi}{10.1016/0022-5096(92)90020-3}}.

\bibitem[{Chen et~al.(2001)Chen, Fleck, and Lu}]{Chen2001}
\bibinfo{author}{C.~Chen}, \bibinfo{author}{N.~A. Fleck},
  \bibinfo{author}{T.~J. Lu}, \bibinfo{title}{Mode {I} crack growth resistance
  of metallic foams}, \bibinfo{journal}{Journal of the Mechanics and Physics of
  Solids} \bibinfo{volume}{49}~(\bibinfo{number}{2}) (\bibinfo{year}{2001})
  \bibinfo{pages}{231--259},
  \doi{\bibinfo{doi}{10.1016/S0022-5096(00)00039-9}}.

\bibitem[{Elices et~al.(2002)Elices, Guinea, Gomez, and Planas}]{Elices2002}
\bibinfo{author}{M.~Elices}, \bibinfo{author}{G.~V. Guinea},
  \bibinfo{author}{J.~Gomez}, \bibinfo{author}{J.~Planas}, \bibinfo{title}{The
  cohesive zone model: advantages, limitations and challenges},
  \bibinfo{journal}{Engineering Fracture Mechanics}
  \bibinfo{volume}{69}~(\bibinfo{number}{2}) (\bibinfo{year}{2002})
  \bibinfo{pages}{137 -- 163},
  \doi{\bibinfo{doi}{10.1016/S0013-7944(01)00083-2}}.

\bibitem[{Ashby et~al.(2000)Ashby, Evans, Fleck, Gibson, Hutchinson, and
  Wadley}]{Ashby2000}
\bibinfo{author}{M.~F. Ashby}, \bibinfo{author}{A.~G. Evans},
  \bibinfo{author}{N.~A. Fleck}, \bibinfo{author}{L.~J. Gibson},
  \bibinfo{author}{J.~W. Hutchinson}, \bibinfo{author}{H.~N.~G. Wadley},
  \bibinfo{title}{Metal Foams: A Design Guide}, \bibinfo{publisher}{Elsevier},
  \bibinfo{year}{2000}.

\bibitem[{Smith(1974)}]{Smith1974}
\bibinfo{author}{R.~A. Smith}, \bibinfo{title}{Calibrations for the electrical
  potential method of crack growth measurement by a direct electrical analogy},
  \bibinfo{journal}{Strain} \bibinfo{volume}{10}~(\bibinfo{number}{4})
  (\bibinfo{year}{1974}) \bibinfo{pages}{183--187},
  \doi{\bibinfo{doi}{10.1111/j.1475-1305.1974.tb00113.x}}.

\bibitem[{Amsterdam et~al.(2008)Amsterdam, de~Vries, Hosson, and
  Onck}]{Amsterdam2008}
\bibinfo{author}{E.~Amsterdam}, \bibinfo{author}{J.~H.~B. de~Vries},
  \bibinfo{author}{J.~T.~M.~D. Hosson}, \bibinfo{author}{P.~R. Onck},
  \bibinfo{title}{The influence of strain-induced damage on the mechanical
  response of open-cell aluminum foam}, \bibinfo{journal}{Acta Materialia}
  \bibinfo{volume}{56}~(\bibinfo{number}{3}) (\bibinfo{year}{2008})
  \bibinfo{pages}{609--618},
  \doi{\bibinfo{doi}{10.1016/j.actamat.2007.10.034}}.

\bibitem[{{Otsu}(1979)}]{Otsu1979}
\bibinfo{author}{N.~{Otsu}}, \bibinfo{title}{A Threshold Selection Method from
  Gray-Level Histograms}, \bibinfo{journal}{IEEE Transactions on Systems, Man,
  and Cybernetics} \bibinfo{volume}{9}~(\bibinfo{number}{1})
  (\bibinfo{year}{1979}) \bibinfo{pages}{62--66},
  \doi{\bibinfo{doi}{10.1109/TSMC.1979.4310076}}.

\bibitem[{Gibson and Ashby(1997)}]{Gibson1997}
\bibinfo{author}{L.~J. Gibson}, \bibinfo{author}{M.~F. Ashby},
  \bibinfo{title}{Cellular Solids: Structure and Properties},
  \bibinfo{publisher}{Cambridge University Press}, \bibinfo{year}{1997}.

\bibitem[{Pineau(1982)}]{Pineau1981}
\bibinfo{author}{A.~Pineau}, \bibinfo{title}{Review of fracture micromechanisms
  and a local approach to predicting crack resistance in low strength steels.
  In: Advances in Fracture Research, ICF5 (Edited by François, D.},
  \bibinfo{publisher}{Pergamon, Oxford, UK}, \bibinfo{year}{1982}.

\bibitem[{O'dowd and Shih(1991)}]{Odowd1991}
\bibinfo{author}{N.~P. O'dowd}, \bibinfo{author}{C.~F. Shih},
  \bibinfo{title}{Family of crack-tip fields characterized by a triaxiality
  parameter—I. Structure of fields}, \bibinfo{journal}{Journal of the
  Mechanics and Physics of Solids} \bibinfo{volume}{39}~(\bibinfo{number}{8})
  (\bibinfo{year}{1991}) \bibinfo{pages}{989--1015},
  \doi{\bibinfo{doi}{10.1016/0022-5096(91)90049-T}}.

\bibitem[{O'Dowd and Shih(1992)}]{Odowd1992}
\bibinfo{author}{N.~P. O'Dowd}, \bibinfo{author}{C.~F. Shih},
  \bibinfo{title}{Family of crack-tip fields characterized by a triaxiality
  parameter—II. Fracture applications}, \bibinfo{journal}{Journal of the
  Mechanics and Physics of Solids} \bibinfo{volume}{40}~(\bibinfo{number}{5})
  (\bibinfo{year}{1992}) \bibinfo{pages}{939--963},
  \doi{\bibinfo{doi}{10.1016/0022-5096(92)90057-9}}.

\bibitem[{Xu and Needleman(1994)}]{Xu1994}
\bibinfo{author}{X.-P. Xu}, \bibinfo{author}{A.~Needleman},
  \bibinfo{title}{Numerical simulations of fast crack growth in brittle
  solids}, \bibinfo{journal}{Journal of the Mechanics and Physics of Solids}
  \bibinfo{volume}{42}~(\bibinfo{number}{9}) (\bibinfo{year}{1994})
  \bibinfo{pages}{1397--1434},
  \doi{\bibinfo{doi}{10.1016/0022-5096(94)90003-5}}.

\bibitem[{Combaz et~al.(2011)Combaz, Rossoll, and Mortensen}]{Combaz2011}
\bibinfo{author}{E.~Combaz}, \bibinfo{author}{A.~Rossoll},
  \bibinfo{author}{A.~Mortensen}, \bibinfo{title}{Hole and notch sensitivity of
  aluminium replicated foam}, \bibinfo{journal}{Acta Materialia}
  \bibinfo{volume}{59}~(\bibinfo{number}{2}) (\bibinfo{year}{2011})
  \bibinfo{pages}{572--581},
  \doi{\bibinfo{doi}{10.1016/j.actamat.2010.09.061}}.

\end{thebibliography}
